\documentclass[aps,prc,reprint,amsmath,amssymb,superscriptaddress,showkeys]{revtex4-1}

\usepackage[utf8]{inputenc}
\usepackage{graphicx}
\usepackage{amsmath}
\usepackage{amsfonts}
\usepackage{amsmath}
\usepackage{amssymb}
\usepackage{soul}
\usepackage{url}
\usepackage{bm}
\usepackage{amsthm}
\usepackage{mathtools}
\usepackage{dsfont}
\usepackage{stmaryrd}
\usepackage{adjustbox}
\usepackage{makecell}
\usepackage{epigraph}
\usepackage{cancel}
\usepackage{hyperref}
\usepackage[dvipsnames]{xcolor}

\usepackage{caption}
\DeclareCaptionJustification{plainjust}{\leftskip=0pt\rightskip=0pt\parfillskip=0pt plus 1fil\relax}
\definecolor{brickred}{rgb}{0.8, 0.25, 0.33}
\newcommand\myshade{85}
\colorlet{mylinkcolor}{BrickRed}
\colorlet{mycitecolor}{NavyBlue}
\colorlet{myurlcolor}{Aquamarine}

\hypersetup{
  linkcolor  = mylinkcolor!\myshade!black,
  citecolor  = mycitecolor!\myshade!black,
  urlcolor   = myurlcolor!\myshade!black,
  colorlinks = true,
}

\newcommand{\bs}[1]{\boldsymbol{#1}}

\newcommand{\TC}{\ensuremath{\mathrm{TC}}}
\newcommand{\DTC}{\ensuremath{\mathrm{DTC}}}

\definecolor{darkgreen}{RGB}{50,200,8}

\renewcommand{\bs}[1]{\boldsymbol{#1}}

\graphicspath{ {./figs/} }

\newcommand\figsubref[2]{\hyperref[#1]{\ref*{#1}#2}}

\usepackage{tikz/tikz-network}

\usepackage{pgfplots}
\pgfplotsset{compat = newest}

\usetikzlibrary{shapes.misc, positioning}
\usetikzlibrary{backgrounds,matrix}

\usetikzlibrary{shapes}
\usetikzlibrary{fit}
\usetikzlibrary{chains}
\usetikzlibrary{arrows}

\tikzstyle{latent} = [circle,fill=white,draw=black,inner sep=1pt,
minimum size=15pt, font=\fontsize{10}{10}\selectfont, node distance=1]
\tikzstyle{obs} = [latent,fill=gray!25]
\tikzstyle{const} = [rectangle, inner sep=0pt, node distance=1]
\tikzstyle{factor} = [rectangle, fill=black,minimum size=5pt, inner
sep=0pt, node distance=0.4]
\tikzstyle{det} = [latent, diamond]

\tikzstyle{plate} = [draw, rectangle, rounded corners, fit=#1]
\tikzstyle{wrap} = [inner sep=0pt, fit=#1]
\tikzstyle{gate} = [draw, rectangle, dashed, fit=#1]

\tikzstyle{caption} = [font=\footnotesize, node distance=0] %
\tikzstyle{plate caption} = [caption, node distance=0, inner sep=0pt,
below left=5pt and 0pt of #1.south east] %
\tikzstyle{factor caption} = [caption] %
\tikzstyle{every label} += [caption] %

\usetikzlibrary{decorations.pathreplacing}

\usepgfplotslibrary{fillbetween,groupplots}
\usepgfplotslibrary{colorbrewer}
\usepgfplotslibrary{statistics}

\usetikzlibrary{pgfplots.external}
\usetikzlibrary{external}

\usepackage{filemod}
\newcommand{\includetikz}[2]{%
  \tikzsetnextfilename{#2}%
  \filemodCmp{#1#2.tex}{#1external/#2.pdf}%
    {\tikzset{external/remake next}}{}%
  \input{#1#2.tex}%
}

\newcommand{\includetikzonecol}[3][\textwidth]{%
  \tikzsetnextfilename{#3}%
  \filemodCmp{#2#3.tex}{#2external/#3.pdf}%
    {\tikzset{external/remake next}}{}%
  \resizebox{#1}{!}{\input{#2#3.tex}}%
}

\usepackage{pgfplotstable}
\newcommand\PlotNetwork[3]{%
  \def\oPos{#1}%
  \def\CIJ{#2}%
  \def\lbl{#3}%
  \node[draw=none, fill=none] (ori) at \oPos {};%
  \foreach \i in {0,1,...,5}%
  {%
    \pgfmathsetmacro{\angle}{\i*360/6}%
    \node[minimum size=\nodeR] (\i) at ($(ori) + (\angle:\netR)$) {};%
  }%
  \foreach \i in {0,1,...,5}%
  {%
    \foreach \j in {0,1,...,5}%
    {%
      \pgfmathtruncatemacro\w{\CIJ[\i][\j]}%
      \ifnum\w>\th\draw[conn]  (\j) -- (\i);\fi%
    }%
  }%
  \node[draw=none, fill=none] (l) at ($(ori) + (135:\netR*1.4)$) {\bfseries \lbl};%
}

\pgfplotsset{select coords between index/.style 2 args={
    x filter/.code={
        \ifnum\coordindex<#1\fi
        \ifnum\coordindex>#2\fi
    }
}}

\pgfplotsset{
  /pgfplots/scatter legend/.style={
    /pgfplots/legend image code/.code={\draw[##1,mark size=1.1pt,yshift=-0.1em, xshift=7pt] plot coordinates {
      (-0.1em, 0.1em)
      (0.2em, 0.45em)
      (0.3em, -0.05em)
      (0.6em, 0.3em)
    };},
    rounded corners=1.5pt,
  },
}

\renewcommand{\includetikz}[2]{\includegraphics{#1external/#2.pdf}}
\renewcommand{\includetikzonecol}[3][\textwidth]{\resizebox{#1}{!}{\includegraphics{#2external/#3.pdf}}}

\begin{document}

\title{Statistical Dark Matter: Synergy is everywhere but is hard to capture}

\newcommand{\eqcontr}{\thanks{These authors contributed equally to this work. \\ Corresponding author: 
\href{mailto:a.liardi@imperial.ac.uk}{a.liardi@imperial.ac.uk}}}
\author{Alberto Liardi}\eqcontr
\affiliation{Department of Computing, Imperial College London, London, United Kingdom}

\author{Fernando E. Rosas} \eqcontr
\affiliation{Sussex AI and Sussex Centre for Consciousness Science, Department of Informatics, University of Sussex, Brighton, United Kingdom}
\affiliation{Department of Brain Science, Imperial College London, London, United Kingdom}
\affiliation{Center for Complexity Science, Imperial College London, London, United Kingdom}
\affiliation{Center for Eudaimonia and Human Flourishing, University of Oxford, Oxford, United Kingdom}

\author{Daniele Marinazzo}
\affiliation{Department of General and Applied Psychology, Ghent University, Ghent, Belgium}

\author{Thomas~F.~Varley}
\affiliation{Vermont Complex Systems Institute, University of Vermont, Burlington, VT, USA}

\author{Michael Gastpar}
\affiliation{School of Computer and Communication Sciences, EPFL, Lausanne, Switzerland}

\author{Pedro A.M. Mediano}
\affiliation{Department of Computing, Imperial College London, London, United Kingdom}
\affiliation{Division of Psychology and Language Sciences, University College London, London, United Kingdom}

\begin{abstract}

\noindent
The information-theoretic construct of \textit{synergy} refers to the statistical structure that is contained in three or more variables, but not in any subset of them. 
Here we leverage recent advances in multivariate information theory to show that synergy is far more prevalent in complex systems than previously thought. 
In particular, we show that many-body systems tend to become strongly dominated by synergy as the number of subcomponents grows. 
At the same time, our results also reveal that commonly used tools for statistical modelling severely underestimate synergistic structures. 
These findings imply that synergy constitutes a prevalent, yet often invisible, informational component of complex systems. With a certain poetic licence, we interpret these results as suggesting that \emph{synergy is the dark matter of statistics} --- interdependencies that we know exist, but standard instruments fail to detect. 
\end{abstract}

\maketitle

\section*{Introduction}

Complex systems are characterised by emergent behaviour arising from higher-order interdependencies, which make ``the whole more than the sum of the parts''~\cite{waldrop1993complexity,crutchfield1994calculi,jensen2022complexity,battiston2021physics,battiston2026collective}. 
The assessment of higher-order interdependencies has enabled powerful practical analyses in a wide range of fields, providing important insights into, for example, collective mechanisms in natural evolution~\cite{rajpal2023quantifying}, genetic information flow~\cite{cang2020inferring}, human behaviour~\cite{battiston2025higher} and psychometric questionnaires~\cite{marinazzo2024information,varley2022untangling}, cellular automata~\cite{Rosas2018}, artificial neural networks~\cite{tax2017partial,proca2024synergistic,kaplanis2023learning,gutknecht2025shannon}, and the human brain~\cite{gatica2021high,luppi2022synergistic,herzog2022genuine,varley2023information,varley2023multivariate}.

\begin{figure}[b!]
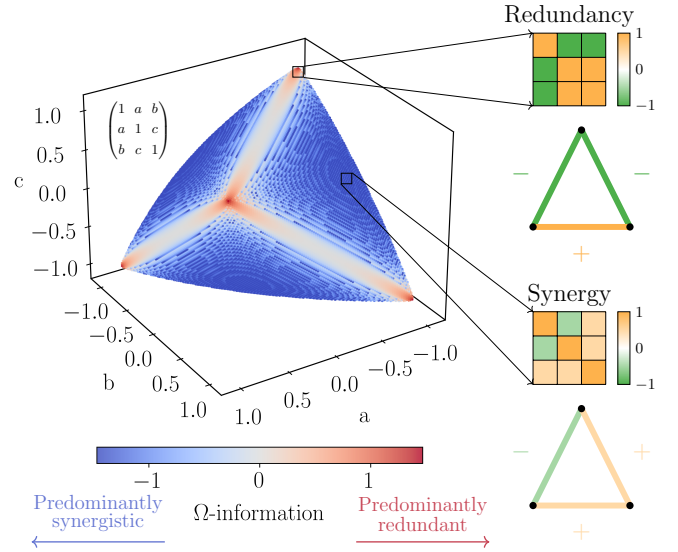

    \centering
    \includetikzonecol[\linewidth]{tikz/}{Fig0}
    \caption{\small\textbf{The informational structure of three-dimensional Gaussian distributions.} Representation of the space of all possible three-dimensional correlation matrices (an elliptope), with the associated O-information and some representative cases highlighted. 
    Negative (blue) and positive (red) O-information indicate the prevalence of synergy and redundancy in the system, respectively. Most of the admissible space corresponds to synergy-dominated systems, with approximately 63\% of the volume characterised by negative O-information.}
    \label{fig:fig0}
\end{figure}

Information theory provides a natural toolbox to study higher-order interdependencies, overcoming limitations of conventional statistical approaches employed in the study of natural systems --- which often rely primarily on two-point correlation functions, pairwise associations, or lower-order summaries~\cite{rosas2022disentangling,varley2023partial}. In particular, \emph{information decomposition}~\cite{williams2010nonnegative, mediano2025toward, ince2017partial,rosas2025characterising, liardi2026mathematical} is a family of approaches that offer robust and quantitative frameworks for dissecting the information shared among a set of variables into qualitatively distinct contributions. 
Among these, \textit{synergy} --- information contained in a group of variables jointly but not in any subset of them --- has proven instrumental in supporting computation in natural~\cite{luppi2024synergistic, luppi2022synergistic, liardi2025null, varley2023multivariate, goodwell2017temporal, varley2022untangling} and artificial~\cite{tax2017partial,proca2024synergistic,gutknecht2025shannon} systems. 
Synergy has further been found to dominate near phase transitions and in systems performing complex distributed information processing~\cite{mediano2022integrated, Rosas2018, marinazzo2019synergy}, and to relate to frustration in spin-like systems~\cite{caprioglio2025synergistic}, technological sophistication in financial markets~\cite{rajpal2025synergistic}, flocking behaviour~\cite{sas2025improved}, non-trivial topological structures~\cite{varley2025topology}, and even dimensions of harmonic complexity in the music of J.S.\ Bach~\cite{scagliarini2022quantifying}.

Although these findings span multiple disciplines, the recurring links between synergy, criticality, and complex computation might suggest that synergistic interdependencies are exceptional phenomena that only arise in specific systems of interest --- akin to critical submanifolds, arising only within a vanishingly small region of the space of possible systems~\cite{hidalgo2014information, mora2011biological}; or complex computation, found only within a narrow region of rule space near the transition between order and chaos~\cite{langton1990computation}. 
In this work, we show that, in large systems, synergy is not the exception but the norm. However, despite its ubiquity, we demonstrate that several of our most widely used statistical models either neglect or substantially underestimate synergistic structure. 
Taken together, our results portray synergy as \textit{the dark matter of statistics}: a pervasive yet largely hidden component of complex systems, present almost everywhere but overlooked by many of our standard observational and inferential tools, which therefore deserves far greater attention in ordinary statistical inquiries.

\section*{Results}

\subsection*{Synergy is prevalent in large systems} 

Although the intuition behind synergy is relatively simple --- namely, statistical dependencies that can only be seen when a collection of components are considered jointly --- providing a principled and general characterisation of it is a formidable challenge~\cite{liardi2026mathematical}.
Fortunately, characterising the balance between synergy (the system's higher-order structure) and redundancy (lower-order dependencies that are shared across components~\cite{bohm2026redundancy}) is substantially easier. 
Recent advances in information theory~\cite{rosas2025characterising} have shown that a unique information-theoretic measure exists that symmetrically balances these informational structures and scales linearly with system size. 
This quantity, known as O-information~\cite{rosas2019quantifying}, is given by
\begin{align}
\Omega(\bs{X}^n) 
& := (n-2) H(\bs{X}^n)
+ \sum_{j=1}^n \big[ H(X_j) -  H(\bs{X}_{-j}) \big] \,,\label{eq:def_o} %
\end{align}
where $\bm X^n = (X_1,\dots,X_n)$ indicates the variables of interest, $\bm X_{-j} =(X_1,\dots,X_{j-1},X_{j+1},\dots,X_n)$, and $H(X)$ is the Shannon entropy of $X$. The O-information can also be expressed in terms of the \emph{Total Correlation} $\TC(\bs{X}^n):=\sum_{j=1}^n H(X_j) - H(\bm X^n)$, a generalisation of Shannon's mutual information~\cite{watanabe1960information, garner1962uncertainty, perez1977varepsilon}, 
as
\begin{equation} \label{eq:def_o2}
    \Omega(\bs{X}^n) = -(n-2)\TC(\bs{X}^n) + \sum_{j=1}^n \TC(\bs{X}_{-j}) \,. 
\end{equation}
Since the total correlation represents global deviation from statistical independence~\cite{watanabe1960information,garner1962uncertainty}, Eq.~\eqref{eq:def_o2} establishes the O-information as a whole-minus-sum comparison between dependency structure at the full-system and subsystem levels.
Accordingly, it has been shown that $\Omega(\bs{X}^n)<0$ characterises systems which are dominated by synergy --- i.e. systems whose synergy is non-zero and, in fact, larger than redundant components~\cite{rosas2019quantifying, rosas2025characterising}. 
Conversely, $\Omega(\bs{X}^n)>0$ characterises systems dominated by redundancy. Due to its computational tractability and interpretational clarity, O-information has found multiple applications~\cite{gatica2021high, gatica2022high, faes2022new, scagliarini2022quantifying, varley2023multivariate}, and negative $\Omega$ is often used as a marker of synergistic information.

\begin{figure}[t]
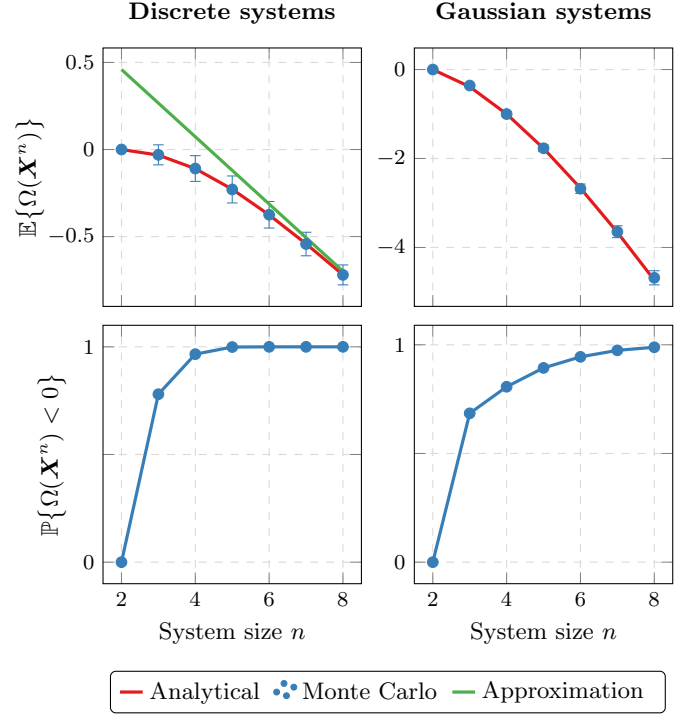

  \centering
  \includetikz{tikz/}{OmegaDirichletWishart}
  \caption{\small\textbf{Large systems are predominantly synergistic}. The average O-information of an ensemble of systems drawn from a uniform Dirichlet distribution (left) or a Wishart distribution (right) tends to $-\infty$ as $n$ grows (top), while the fraction of synergistic systems (i.e. $\Omega<0$) rapidly converges to 1 (bottom). Error bars represent 95\% confidence intervals estimated via Monte Carlo simulation.}
  \label{fig:dirichlet}
\end{figure}

Using the O-information, we first investigate the degree of synergy present in large multivariate systems. %
Our approach is to study families of systems defined via a ``prior'' over specific spaces of probability distributions, and then compute their expected O-information as a function of the system size $n$. 

To investigate systems with discrete variables, we consider arbitrary systems whose state is described by a binary vector $\bs{X}^n\in\{0,1\}^n$. For generality, we use a symmetric Dirichlet distribution $\mathrm{Dir}(2^n, \alpha)$ with dispersion parameter $\alpha$ as a prior over the $(2^n-1)$-dimensional simplex, covering all possible distributions $p(\bs{X}^n)$. 
Symmetric Dirichlet distributions are a standard choice for assessing the hyperparameters of binary systems; in particular, $\alpha=1$ corresponds to a uniform distribution over the $2^n$ probability simplex, and other values of $\alpha$ characterise ensembles with various degrees of dispersion~\cite{Bishop2006}.

\begin{figure*}[ht]
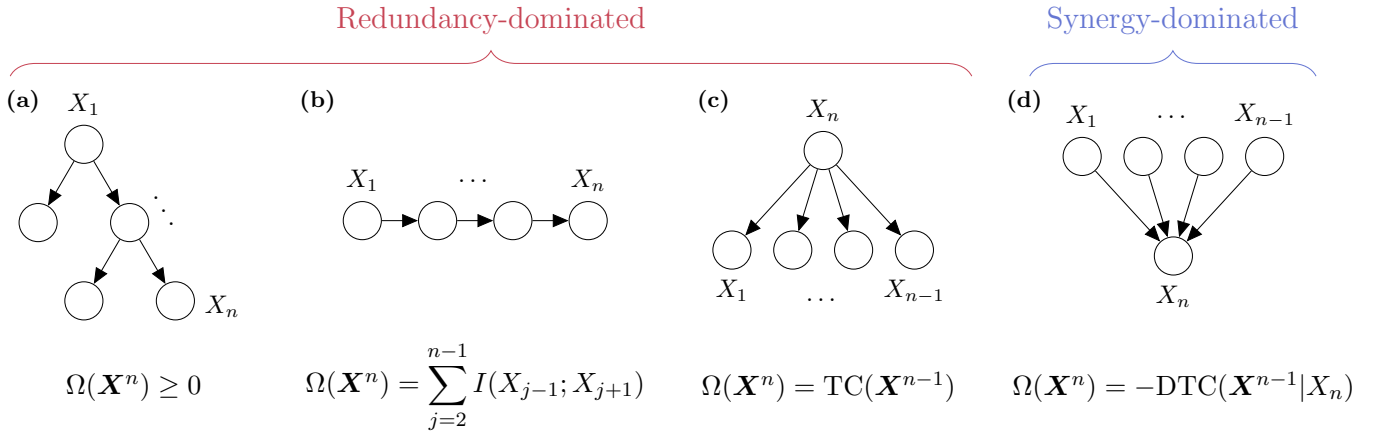

  \centering
  \includetikzonecol[520pt]{tikz/}{BayesNets_v2}
  \caption{\small\textbf{Tree-structured Bayesian networks are redundancy-dominated}.
  General tree-structured graphical models \textbf{(a, b, c)} have been objects of deep study in machine learning. Based on their structure alone, we know they can only represent redundancy-dominated distributions. For some tree-structured models of interest, such as \textbf{(a)} Markov chains and \textbf{(b)} Naive Bayes, we can provide closed-form expressions for $\Omega$. Note that other graphical models, such as \textbf{(c)} a stochastic function of independent inputs, can capture synergy-dominated distributions.} 
  \label{fig:bayesnets}
\end{figure*}

Our results show that the average O-information of binary systems drawn from a $\text{Dir}(2^n,\alpha)$ distribution is
\begin{align}
  \mathbb{E}\big[ \Omega(\bs{X}^n) \big] = \, & (n-2)\big[ \psi(2^n\alpha+1) - \psi(\alpha+1) \big] \nonumber \\
&+ n \big[ \psi(2\alpha+1) - \psi(2^{n-1} \alpha + 1) \big],
\label{eq:dirichlet_mean}
\end{align}
where $\psi$ is the standard digamma function (for technical details, see \textit{Methods}). For large $n$, the expected O-information can be expanded linearly as 
\begin{equation}
\mathbb{E} \big[\Omega(\bs{X}^n) \big] = a \cdot n + b +\mathcal{O}(2^{-n})\,.
\end{equation}    
Crucially, we show that $a<0 \,\,\, \forall \alpha\in \mathbb{R}^+$ (see \textit{Methods}), and hence
\begin{equation}
\lim_{n \rightarrow \infty} \mathbb{E}[\Omega(\bs{X}^n)] = - \infty \,, \label{eq:asymptote_dir}
\end{equation}
providing a formal proof that the average O-information becomes infinitely negative as the system size grows.

We now follow the same strategy for continuous variables, considering zero-mean multivariate Gaussian distributions (Fig.~\ref{fig:fig0}). A natural prior for this case is to parametrise the covariance matrices via a Wishart distribution~\cite{Bishop2006}. In this scenario, our findings show that the expected O-information is (see \textit{Methods} for details)
\begin{align*}
  \mathbb{E} \big[ \Omega(\bm X^n) \big] = & \frac{n}{2}\left[ \psi\left(\frac{n}{2}\right) + \psi\left(\frac{1}{2}\right) \right] - \sum_{i=1}^n \psi\left(\frac{i}{2}\right) ~ .
\end{align*}
As in the discrete case above, this also satisfies $\lim_{n\to\infty}\mathbb{E}[\Omega(\bm X^n)]=-\infty$, diverging as $\mathcal{O}(n \log n)$. 
This result generalises previous reports~\cite{barrett2015exploration} showing that synergy is prevalent in trivariate Gaussian systems.

A potential concern is that a mean value diverging to $-\infty$ could still allow a wide range of outcomes if the variance diverges as well. However, Monte Carlo simulations show that the variance remains relatively stable (Fig.~\ref{fig:dirichlet}), implying that $\Omega(\boldsymbol{X}^n) \ll 0$ for most distributions as $n$ grows. Notably, one needs more than $10^3$ realisations to find one Gaussian system with $\Omega > 0$ for $n=8$. This indicates that the dominance of synergy is not an average driven by rare extreme configurations, but rather a typical property of high-dimensional systems.

\subsection*{Common Bayesian networks are redundancy-dominated}

After showing that systems become increasingly synergistic as their size grows, we now investigate the synergy in structured graphical models~\cite{koller2009probabilistic}, tools used ubiquitously across disciplines to make sense of data. For this purpose, let us first consider graphical models with a tree structure, where each variable $X_i$ has at most one parent variable. The distribution of those systems can be written as
\begin{equation}
p_{\bs{X}^n}(\bs{x}^n) = \prod_{i=1}^n p_{X_i|X_{\pi(i)}}(x_i|x_{\pi (i)} ) ~ ,
\end{equation}
where $X_{\pi (i)}$ is the parent variable of $X_i$. 
Tree-structured graphical models play an important role in the machine learning literature, as they allow exact inference via efficient algorithms~\cite{pearl1982reverend}. Non-tree-structured models are also widely used, but inference is then approximate~\cite{minka2001expectation}.

We show that tree-structured graphical models are redundancy-dominated --- i.e.\, ${\Omega(\bs{X}^n)\geq 0}$ for all members of this class of systems (Fig.~\ref{fig:bayesnets}). This generalises previously established results in the literature for more restricted classes of systems~\cite{rosas2024characterising}. We report all the proofs of the following findings in \textit{Methods}. 

To gain insight into this result, we first consider Markov chains, a class of tree-structured graphical models in which conditioning on an intermediate variable renders the variables on either side conditionally independent. This suggests that such systems are dominated by redundant interactions. The following result confirms this intuition, providing a non-negative closed-form expression for $\Omega$:
\begin{equation}
\Omega(\bs{X}^n) = \sum_{j=2}^{n-1} I(X_{j-1};X_{j+1}) \geq 0 ~ .
\end{equation}

As a second illustrative example, we consider a different type of tree-structured graphical model: \textit{Naive Bayes} distributions. By direct calculation, we obtain that the O-information of these models is given by
\begin{equation}
\Omega(\bs{X}^n) = \sum_{i=1}^{n-1} H(X_i) - H(\bs{X}^{n-1}) = \TC(\bm X^{n-1}) \geq 0 ~ .
\end{equation}
This finding clarifies that the redundancies arise from interdependencies within $\bm X^{n-1}$ that are necessarily mediated by the common parent $X_n$.

Importantly, not all graphical models are redundancy-dominated --- specifically, graphical models can have negative O-information if variables have multiple parents. For instance, consider graphical models in which $X_1,\dots,X_{n-1}$ are jointly independent, and $X_{n}=F(X_1,\dots,X_{n-1})$ with $F(\cdot)$ a (possibly random) function, so that their joint distribution is
\begin{equation}
p_{\bs{X}^n}(\bs{x}) = p_{X_n|\bs{X}^{n-1}}(x_n|\bs{x}^{n-1}) \prod_{i=1}^{n-1} p_{X_i}(x_i) ~ .
\label{eq:indep}
\end{equation}
The O-information of such graphical models can be shown to be (see \textit{Methods}):
\begin{equation}
    \Omega(\bs{X}^n) = - \text{DTC}(\bs{X}^{n-1}|X_n)\leq 0 ~ ,
\end{equation}
where ${\text{DTC}(\bs{X}^n):= H(\bs{X}^n) - \sum_{j=1}^n H(X_j|\bs{X}_{-j})}$ is another non-negative generalisation of Shannon's mutual information known as \emph{Dual Total Correlation}~\cite{han1975linear}.
These models exemplify the operation of colliders~\cite{pearl2009causal}, i.e.\ variables with two or more parents. Here the parents share no dependencies among themselves, so the structure they jointly impose on the child cannot be recovered from any subset of them, giving rise to synergy.
Inferring such structures is computationally demanding: because parents must be handled jointly rather than one at a time, learning general graphical structure from data is NP-hard~\cite{chickering1996learning}, whereas the optimal tree can be recovered in polynomial time~\cite{chow1968approximating}. Accordingly, although many real systems involve multiple converging causes~\cite{berkson1946, munafo2018collider, griffith2020collider}, tree-based approaches are routinely favoured for their tractability, leading to a systematic underestimation of synergistic structure in fitted models. We demonstrate these shortcomings explicitly in the next section. 

\begin{figure*}[ht]
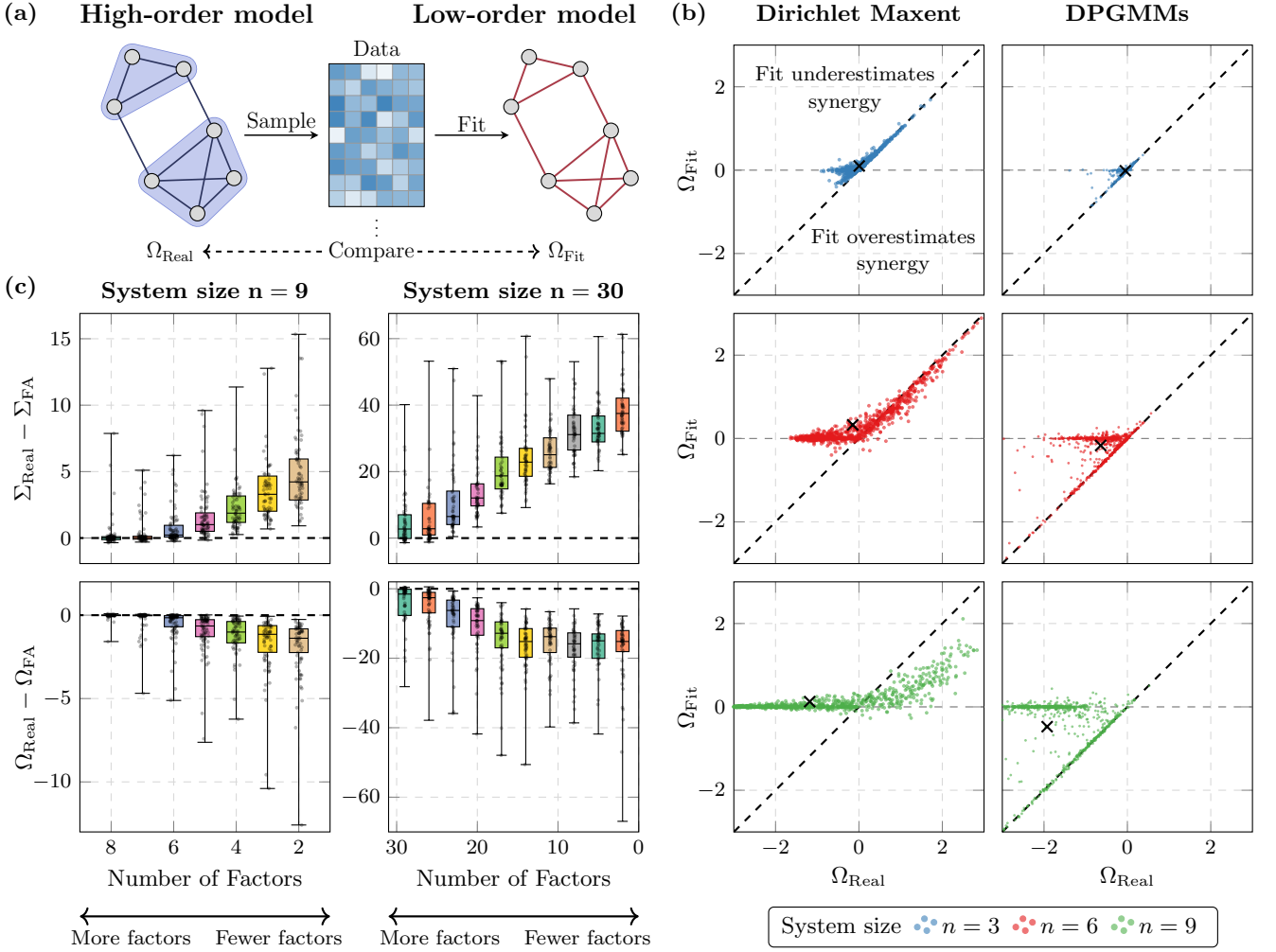

    \centering
    \includetikzonecol[500pt]{tikz/}{LowVsHighOrder}
    \caption{\small\textbf{Common statistical models and factor analyses underestimate synergy.} \textbf{(a)} Schematic of our approach: we sample data from a higher-order model -- i.e.\ a distribution with interactions between three or more variables (shaded) -- then fit a lower-order model with pairwise interactions only, and finally compare the O-information of the original and the fitted distributions. We use Dirichlet distributions and DPGMMs as higher-order models, and Ising and multivariate Gaussian distributions as their respective lower-order fits.
    \textbf{(b)} O-information of Dirichlet distributions and DPGMMs and of their fit with a maximum entropy (Ising) and multivariate Gaussian distribution, respectively, for $n=3,6,9$. 
    \textbf{(c)} Differences between the S- and O-information of DPGMMs and their fit according to a given number of factors, for $n=9$ (left) and $n=30$ (right) variables. 
    }
    \label{fig:synergyloss}
\end{figure*}

\subsection*{Lower-order models underestimate synergy}

The results above show that synergy is widespread across broad families of probability distributions, both in discrete and continuous domains, and that commonly used probabilistic graphical models cannot express such synergistic structures. 
We now demonstrate that this limitation has concrete practical consequences: when data is fitted by a lower-order model, synergy is systematically underestimated --- so the analysis fails to account for key higher-order statistical structure.

To study this, we followed a three-step process: (i) generate data by sampling from a broad family of probability distributions with higher-order structure, (ii) fit the data to a lower-order model, and (iii) compare the O-information of the original and fitted distributions (Fig.~\figsubref{fig:synergyloss}{a}). 
For the first step, we used a symmetric Dirichlet distribution for the discrete case (as above), and a Dirichlet Process Gaussian Mixture Model (DPGMM)~\footnote{DPGMMs can be informally understood as a mixture of Gaussian components with a flexible (potentially unbounded) number of components.} for the continuous case~\cite{neal2000markov}.  
As lower-order models, we used maximum-entropy models under pairwise constraints --- corresponding to Ising (discrete) and Gaussian (continuous) distributions. Note that these models only contain pairwise interactions, but can still exhibit some degree of synergy~\cite{rosas2016understanding,caprioglio2025synergistic}.

Our analyses show that the O-information of the fitted pairwise models is consistently higher than that of the original distributions for both discrete and continuous cases (Fig.~\figsubref{fig:synergyloss}{b}). 
However, since the O-information quantifies the balance between redundancy and synergy, this increase alone cannot discriminate between two scenarios: it could arise either from an increase in redundancy or a loss of synergy. To disentangle these possibilities, we additionally compute the S-information $\Sigma$~\cite{rosas2019quantifying}, which captures the total statistical structure of the system by aggregating both redundant and synergistic contributions (see \textit{Methods}). 
Results show that S-information is consistently underestimated in the lower-order models for both discrete and continuous scenarios (Fig.~\ref{fig:synloss_Sinfo}, \textit{Appendix}~\ref{app:synloss_sinfo}). Together with the increase in the O-information, such a decrease in S-information strongly suggests that pairwise models systematically underestimate synergy. Interestingly, this effect becomes more pronounced as the number of variables increases.
Moreover, this reduction in synergy cannot be attributed to a poor goodness of fit, since the fitted models accurately reproduce the original distribution's first- and second-order statistics, with each entry of the mean vector and covariance matrix being on average within 1\% of its original value. Therefore, the loss of synergy truly reflects the inability of pairwise models to capture higher-order dependencies, and not a failure of parameter estimation.

To further investigate this question, we also studied to what degree factor analysis --- a popular pairwise technique in psychometric analysis~\cite{comrey2013first} --- can capture synergy. For this, we generated samples from a DPGMM and performed confirmatory factor analysis with varying numbers of factors. From the resulting factor models, we then reconstruct covariance matrices and compute the resulting O-information. 

Our analyses show that reducing the number of factors leads to a systematic increase in the estimated $\Omega_{\text{FA}}$, indicating a relative increase in redundancy-dominance --- with a larger increase for bigger systems (Fig.~\figsubref{fig:synergyloss}{c}).
As above, we repeat the same analysis with S-information, obtaining that $\Sigma_{\text{FA}}$ decreases as the number of factors is reduced, again with a stronger effect in larger systems. These findings are not an artefact of fitting a Gaussian model to data generated from a DPGMM --- in fact, an analogous behaviour is observed when factor analysis is applied to covariance matrices sampled from a Wishart distribution (Fig.~\ref{fig:synergyloss_gauss}, Appendix~\ref{app:FA_gaussian}).
Taken together, the opposing trends of O- and S-information indicate that the observed increase in $\Omega_{\text{FA}}$ is driven by a reduction in synergy rather than an increase in redundancy. In other words, the factor models progressively fail to capture higher-order structures as dimensionality is reduced.

\vspace*{-4pt}
\section*{Discussion}

\begin{quotation} 
\noindent\textit{The Tao gave birth to One. One gave birth to Two. Two gave birth to Three. Three gave birth to all things.} 
\end{quotation} 
\begin{flushright} 
Lao Tse, \textit{Tao Te Ching}
\end{flushright}

Synergy represents the statistical structure that becomes available when variables are considered jointly, and that no subset of them can reveal.
Its emergence requires at least three components: a single variable has no interdependencies to speak of, and any dependency between two is, by construction, already visible in the pair. Only with the third does a qualitatively new kind of structure become possible.
Beyond three, no further kind of interdependency appears --- only more of it --- and, as we have shown, overwhelmingly more.

In this work, we used formal and empirical methods to demonstrate that synergy tends to be the dominant informational structure in large complex systems --- both in discrete and Gaussian settings. 
Furthermore, we showed that standard statistical models often used for data analysis and machine learning fail to capture higher-order interdependencies. 
These findings suggest that synergy is ``here, there, and everywhere'', but many of our statistical tools systematically underestimate its presence.

\subsection*{Synergy and real-world scenarios}

Given the links between synergy and emergence~\cite{Rosas2018, mediano2022integrated}, self-organisation~\cite{marinazzo2019synergy, caprioglio2025synergistic}, and collective behaviour in complex systems~\cite{rajpal2023quantifying, battiston2025higher, sas2025improved}, misdetecting synergy may lead to a systematic underestimation of important features in experimental analyses. %

In this work, we have shown that synergy is prevalent in the space of possible probability distributions. Determining whether synergy is also prevalent in real-world systems is, however, a much more challenging (and interesting) question. Answering this question is complicated by the fact that a wide range of physical and biological factors can bias systems towards one or the other end of the redundancy-synergy spectrum. 
For example, although synergy is fundamental in the functioning of complex biological systems~\cite{luppi2022synergistic,chan2017gene}, biological evolution might increase the amount of redundancy by selecting systems that are more robust to internal errors and environmental noise~\cite{krakauer2002redundancy, battail2007information, kitano2007towards}.
In physics, while critical manifolds are often a very small part of a system's parameter space~\cite{mora2011biological}, phenomena like self-organised criticality~\cite{jensen1998self} can push systems towards criticality spontaneously, thus increasing synergy.
Hence, it is currently unknown whether complex large systems are as unbalanced towards synergy as random ones, and therefore to what extent they suffer from the modelling shortcomings above. 

Even though a definitive answer to this question remains distant, further exploration of how prevalent synergy is in real-world systems and what factors are responsible for it remains a promising avenue for future work.

\subsection*{Limitations and future work}

In this work, we provided a suite of mathematical results characterising the presence of synergy in commonly used statistical and machine learning models. As such, our analyses concern synthetic and random data rather than real-world systems.
Therefore, a promising avenue of future work consists of measuring how much synergistic structure is lost when lower-order models are fitted to empirical datasets, across domains and system sizes, to establish how pervasive this underestimation is in practice.
Achieving so requires methods to efficiently estimate synergy from data, on which work is also ongoing~\cite{liardi2025scalable,ehrlich2024partial,bounoua2024somegai}.

Moreover, it is worth noting that the O-information measures the balance between redundancy and synergy. Consequently, a decrease in an already-negative O-information can reflect an increase in synergy, a decrease in redundancy, or a combination of both. Here we mitigated this shortcoming by additionally calculating S-information. 
Various information-theoretic approaches (such as Partial Information Decomposition~\cite{williams2010nonnegative,liardi2026mathematical} or Partial Entropy Decomposition~\cite{ince2017partial}) have been recently introduced to resolve this ambiguity by providing a direct estimation of redundancy and synergy separately. Future work continuing the theory and application of these frameworks could help settle these questions more decisively.

Finally, our results underscore the need for statistical models able to represent higher-order dependencies at scale.
Considerable progress has been made on this line of work, from inferring hypergraphs and simplicial complexes from data~\cite{young2021hypergraph,malizia2024reconstructing}, to Bayesian approaches to hyperedge inference~\cite{contisciani2022inference} or maximum-entropy models beyond pairwise couplings~\cite{ganmor2011sparse, cayco2018moment, decelle2025inferring}.
Future research should focus on making these approaches tractable at scale. In practice, many current techniques target interactions of triplets or quadruplets, whereas our results indicate that synergistic structure becomes overwhelmingly more prominent as larger sets of variables are considered. Closing this gap would make synergy routinely accessible to empirical analysis.

\subsection*{Conclusion}

Synergistic, higher-order dependencies are crucial for describing the collective phenomena that arise in complex systems. Our study shows that these structures are dominant across the great majority of possible systems, yet current statistical tools fail to capture them adequately. 
Thus, this work underscores the need for new analytical methods that can accurately quantify higher-order interdependencies. Achieving this is essential for advancing our understanding of emergence, self-organisation, and collective behaviour in large interactive systems. 

We hope that these findings encourage the scientific community to develop and adopt more sophisticated tools, enabling researchers to uncover the rich informational structure that underlies the complexity of the natural and artificial systems under examination.

\section*{Methods} \label{sec:methods}
\subsection*{Expected $\Omega$ in discrete systems}
\label{app:dirichlet}

In this section, we derive the expression for the expected value of $\Omega(\bs{X}^n)$ when $p(\bs{X}^n)$ is drawn from a symmetric Dirichlet distribution with concentration parameter $\alpha$. Let us define $\mathrm{Dir}(K, \alpha)$ to be the Dirichlet distribution over the $(K-1)$ simplex with all parameters $\alpha_1 = \dots = \alpha_K =\alpha$. 

Let us start by considering a binary distribution $p(\bs{X}^n) \sim \mathrm{Dir}(2^n, \alpha)$.  From this joint distribution, one can calculate the individual marginals by adding all $2^n$ components of $p(\bs{X}^n)$ into two numbers. By using the aggregation property of the Dirichlet distribution, one can find that
\begin{align}
  p(X_j) \sim \mathrm{Dir}(2, 2^{n-1}\alpha) ~ .
\end{align}
Similarly, $p(\bs{X}_{-j}^n)$ is obtained by adding pairs of numbers and results in a distribution with $2^{n-1}$ components. Hence, it is distributed as
\begin{align}
  p(\bs{X}_{-j}^n) \sim \mathrm{Dir}(2^{n-1}, 2\alpha) ~ .
\end{align}

With these results, one can find the mean value of $\Omega(\bs{X}^n)$ by rewriting Eq.~\eqref{eq:def_o} in terms of expected joint entropies:
\begin{multline}  
  \mathbb{E}[\Omega(\bs{X}^n)] = (n-2) \mathbb{E}\left[H(\bs{X}^n)\right] \\ + n\left(\mathbb{E}[H(X_j)] -  \mathbb{E}[H(\bs{X}_{-j})]\right) ~ .
  \label{eq:o_entropy}
\end{multline}
Note that this equality is a consequence of the linearity of $\mathbb{E}[\cdot]$. Results presented by Nemenman \emph{et al.}~\cite{nemenman2002entropy} show that the expected value of the entropy of a distribution with $K$ components drawn from a symmetric Dirichlet with parameter $\alpha$ is given by
\begin{align}
  \mathbb{E}\left[H|K, \alpha\right] = \psi(K\alpha + 1) - \psi(\alpha + 1) ~ ,
  \label{eq:H_dirichlet}
\end{align}
\noindent where $\psi(\cdot)$ is the standard digamma function, and the entropy is measured using natural logarithms. Applying this result to our scenario yields
\begin{subequations}
\begin{align}
  \mathbb{E}[H(X_j)]           &= \psi(2^n\alpha + 1) - \psi(2^{n-1}\alpha + 1) ~ , \\
  \mathbb{E}[H(\bs{X}_{-j})] &= \psi(2^n\alpha + 1) - \psi(2\alpha + 1) ~ , \\
  \mathbb{E}[H(\bs{X}^n)]      &= \psi(2^n\alpha + 1) - \psi(\alpha + 1) ~ .
\end{align}
\end{subequations}
A direct calculation using these identities on Eq.~\eqref{eq:o_entropy} arrives at Eq.~\eqref{eq:dirichlet_mean}, which we rewrite here for convenience:
\begin{align}
  \mathbb{E}\big[ \Omega(\bs{X}^n) \big] = \, & (n-2)\big[ \psi(2^n\alpha+1) - \psi(\alpha+1) \big] \nonumber \\
&+ n \big[ \psi(2\alpha+1) - \psi(2^{n-1} \alpha + 1) \big] \,.
\label{eq:dirichlet_mean_methods}
\end{align}

Now we proceed to prove that this expected $\Omega(\bm X^n)$ diverges to negative infinity as the number of bits grows. We will do this in two steps: first, we show that in the large $n$ limit it grows linearly with $n$, and then we show that the slope of this line is always negative. We start by writing the expected O-information for large $n$ as
\begin{align}
\begin{split}
  \mathbb{E} \big[\Omega(\bs{X}^n) \big] = a \cdot n + b +
\mathcal{O}(2^{-n})
  \label{eq:omega_slope}%
\end{split}
\end{align}

For finding the coefficients of the linear approximation given in Eq.~\eqref{eq:omega_slope}, let us assume that $2^{n-1}\alpha \gg 1$ and use the asymptotic approximation $\psi(x+1) \approx \ln x - \mathcal{O}(x^{-1})$. With this, one can rewrite the first and fourth terms in Eq.~\eqref{eq:dirichlet_mean_methods} as
\begin{equation} \label{eq:linear_approx}
\begin{split}
(n-2)\psi(2^n\alpha + 1) -& n\psi(2^{n-1}\alpha + 1) = \\
  &= (n-2)\ln(2^n\alpha) - n\ln(2^{n-1}\alpha) \\
  &= - n \ln 2 - 2 \ln \alpha
~ .
\end{split}
\end{equation}
The functional form of the coefficients is then obtained from Eqs.~\eqref{eq:dirichlet_mean_methods} and Eq.~\eqref{eq:linear_approx} by grouping the terms that are linear in $n$, yielding
\begin{subequations}
\begin{align} \label{eq:a_coef}
  a &= \psi(2\alpha+1) - \psi(\alpha+1) - \ln 2, \\
  b &= 2 \psi(\alpha+1) - 2\ln \alpha .
\end{align}
\end{subequations}

Finally, let us show that $a$ (Eq.~\eqref{eq:a_coef}) is negative for all $\alpha\in(0,\infty)$. For this, we use the bounds
\begin{align*}
  \ln(x-\frac{1}{2}) < \psi(x) < \ln x ~,
\end{align*}
which hold for all $x\in (1/2, \infty)$. Using these bounds on Eq.~\eqref{eq:a_coef} one finds that
\begin{align*}
  a  & < \ln(2\alpha+1) - \ln(\alpha+1/2) - \ln 2 \\
  & = \ln 2 + \ln(\alpha+1/2) - \ln(\alpha+1/2) - \ln 2 \\
  & = 0 ~ .
\end{align*}

As an alternative proof, note that by comparing Eq.~\eqref{eq:a_coef} with Eq.~\eqref{eq:H_dirichlet}, the difference of digammas corresponds to the expected entropy of a bit drawn from a $\mathrm{Dir}(2,\alpha)$ distribution. Therefore, this difference is bounded between $0$ and $\ln 2$, arriving at the same result as above.

\subsection*{Expected $\Omega$ in Gaussian systems}
\label{sec:wishart}

Now we do a similar calculation as above, but for Gaussian variables. Since the O-information does not depend on the means, to define a distribution over Gaussians, we only need to define a distribution over covariance matrices. Due to its widespread use, we choose the Wishart distribution. Therefore, we want to calculate $\mathbb{E}[\Omega(\Sigma)]$, with
\begin{align}
  \Sigma \sim \mathcal{W}(S, \nu) ~ ,
\end{align}
where $S$ is the \emph{base matrix} and $\nu$ the \emph{degrees of freedom} of the Wishart distribution. The symbol $n$ denotes the number of dimensions of the Gaussian, and we will take as the base matrix the identity matrix, $S = \bm I_n$.

We will use the usual formula for the entropy of a multivariate normal:
\begin{align}
  H(\Sigma) = \frac{1}{2} \log |2\pi e \Sigma |
\end{align}

The expression of $\Omega$ is
\begin{align*}
  \Omega(\bm X^n) &= \TC(\bm X^n) - \DTC(\bm X^n) \\
    &= \sum_{j=1}^n H(X_j) - H(\bm X^n) - \left( H(\bm X^n) - \sum_{j=1}^n R_j \right)
\end{align*}
And using the additivity of the expectation operator, we have
\begin{align*}
  \mathbb{E}[\Omega(\bm X^n)] = -2 \mathbb{E}[H(\bm X^n)] + n( \mathbb{E}[H(X_i)] + \mathbb{E}[R_i] )
\end{align*}

The joint entropy can be easily obtained from~\cite[Appendix A.4, Proposition
2]{Gupta2010}, and yields:
\begin{align}
  \mathbb{E}[H(\bm X^n)] &= \frac{n}{2} \log 2\pi e + \frac{1}{2} \log |2 \bm I_n| + \frac{1}{2} \sum_{i=1}^n \psi(\frac{\nu - n + i}{2}) \nonumber \\
  &= \frac{n}{2} \log 4\pi e + \frac{1}{2} \sum_{i=1}^n \psi(\frac{\nu - n + i}{2})
\end{align}

To calculate $\mathbb{E}[H(X_i)]$, we will use the Bartlett decomposition of $\Sigma$, which says that
\begin{align*}
  \Sigma = L A A^T L^T ~ ,
\end{align*}
where $L$ is the Cholesky factor of $S$ (in our case, $L= \bm I_n$), and $A$ is a $n$-by-$n$ matrix given by
\begin{align*}
  A = \begin{pmatrix}   c_1 &      0 &   0 & ~      \\
                     n_{12} &    c_2 &   0 & \cdots \\
                     n_{13} & n_{23} & c_3 & ~      \\
                          ~ & \vdots &   ~ & \ddots \\ \end{pmatrix}
\end{align*}
where $c_i^2 \sim \chi_{\nu-i+1}^2$ and $n_{ij} \sim \mathcal{N}(0,1)$
independently~\cite[p. 257]{Anderson2003}. A direct calculation of $\Sigma = A
A^T$ in our case yields
\begin{align*}
  \Sigma_{ii} = c_i^2 + \sum_{j=1}^{i-1} n_{ji}^2 \sim \chi_{\nu}^2 ~ .
\end{align*}

The last step can be easily verified using the fact that $\chi^2_k$ is defined as the sum of $k$ squared random standard-normal numbers. To this result, we have to add the expectation of the logarithm of a $\chi_k^2$ number: if $y \sim \chi_k^2$, then~\cite[Abstract]{Pav2015}
\begin{align*}
  \mathbb{E}[\log y] = \log 2 + \psi\left(\frac{k}{2}\right)
\end{align*}
With these results in hand, we can calculate the entropy of the marginals
\begin{align}
  \mathbb{E}[H(X_i)] &= \frac{1}{2}\log 2\pi e \Sigma_{ii} \nonumber \\
  &= \frac{1}{2}\log 4 \pi e + \frac{1}{2}\psi\left(\frac{\nu}{2}\right) ~ .
\end{align}

We now move to the residuals, $R_i$. This is also the entropy of a univariate Gaussian, but this time conditioned on all other variables. Let us define $\bm\delta = \{1, \dots i-1, i+1, \dots n\}$. Then the variance of this Gaussian is
\begin{align*}
  \rho_i = \Sigma_{ii} - \Sigma_{i \bm\delta} \Sigma_{\bm\delta \bm\delta}^{-1} \Sigma_{\bm\delta i}
\end{align*}
This is known as the \emph{scalar Schur complement}, and conveniently it is also $\chi^2$-distributed~\cite[Sec. 4, par. 1]{Siriteanu2016}:
\begin{align*}
  \rho_i \sim \chi_{\nu-n+1}^2 ~ .
\end{align*}
Now, using the same log-expectation formula as above, we arrive at
\begin{align}
  \mathbb{E}[R_i] &= \frac{1}{2}\log 2\pi e \rho_i \nonumber \\
  &= \frac{1}{2}\log 4 \pi e + \frac{1}{2}\psi\left(\frac{\nu-n+1}{2}\right) ~ .
\end{align}

Putting all of these together and simplifying, we arrive at the main result
\begin{align}
  \mathbb{E}[\Omega(\bm X^n)] = & \frac{n}{2}\left[ \psi\left(\frac{\nu}{2}\right) + \psi\left(\frac{\nu-n+1}{2}\right) \right] \nonumber \\
  & - \sum_{i=1}^n \psi\left(\frac{\nu-n+i}{2}\right)
  \label{eq:o_wish}
\end{align}

Matrices sampled from a Wishart distribution are only valid full-rank covariance matrices (i.e.\ symmetric positive definite) if $\nu \geq n$, so we cannot simply take the $n \rightarrow \infty$ limit in Eq.~\eqref{eq:o_wish} with a fixed $\nu$. Instead, we set $\nu = n$ to get the following expression:
\begin{align*}
  \mathbb{E}[\Omega(\bm X^n)] = & \frac{n}{2}\left[ \psi\left(\frac{n}{2}\right) + \psi\left(\frac{1}{2}\right) \right] - \sum_{i=1}^n \psi\left(\frac{i}{2}\right) ~ .
\end{align*}

We are interested in the $n \rightarrow \infty$ limit of the equation above. The only term getting in our way is the sum of digamma functions in the last term. Since we are only interested in the large-$n$ behaviour, we can consider even $n$ and break this term into odd and even $i$:
\begin{align*}
  \sum_{i=1}^n \psi\left(\frac{i}{2}\right) = \sum_{i=1}^{n/2} \psi\left(i\right) \; + \; \sum_{i=1}^{n/2} \psi\left(i - \frac{1}{2}\right) ~ .
\end{align*}
Now that the sums advance in integer steps inside the $\psi(\cdot)$, we can make use of the digamma's recurrence relation, $\psi(x+1) - \psi(x) = 1/x$, to provide a closed form:
\begin{alignat*}{2}
  &\sum_{i=1}^{n/2} \psi\left(i - \frac{1}{2}\right) && = \frac{1}{2}\left[(n-1) \psi\left(\frac{n+1}{2}\right) + \psi\left(\frac{1}{2}\right) - n \right] ~ , \\
  &\sum_{i=1}^{n/2} \psi\left(i\right) && = \frac{n}{2} \left[ \psi\left(\frac{n}{2}+1\right) - 1 \right] ~ .
\end{alignat*}
Combining the equations above and simplifying, we arrive at
\begin{align*}
  \mathbb{E}[\Omega(\bm X^n)] =& -1 -\frac{1}{2}\psi\left(\frac{1}{2}\right) + n \left[1 + \frac{1}{2}\psi\left(\frac{1}{2}\right)\right] \\ &- \frac{n-1}{2}\psi\left(\frac{n+1}{2}\right) ~ .
\end{align*}
From this equation, the last term is the fastest-growing with $\mathcal{O}(n\log n)$, and because of its negative sign, we conclude that
\begin{align*}
  \lim_{n \rightarrow \infty} \mathbb{E}[\Omega(\bs{X}^n)] = - \infty ~ .
\end{align*}

The proof for odd $n$ proceeds analogously and arrives at the same result that random Gaussian systems are predominantly synergistic in the large-$n$ limit.

\subsection*{O-information in probabilistic graphical models}

\subsubsection*{Tree-structured graphical models are redundancy-dominated}
\label{app:graphical}

To begin, recall that the total correlation, dual total correlation, and O-information can conveniently be written as (\cite{rosas2019quantifying} Cor.\@1):
\begin{align}
\TC(\bs{X}^n) &= \sum_{i=2}^n I(X_i;\bs{X}^{i-1})~,\label{bound_TC} \\
\DTC(\bs{X}^n) &= I(X_n;\bs{X}^{n-1}) + \sum_{j=2}^{n-1} I(X_j;\bs{X}^{j-1}|\bs{X}_{j+1}^n), \label{bound_DTC}\\
\Omega(\bs{X}^n) &= \sum_{k=2}^{n-1} I(X_k;\bs{X}^{k-1} ; \bs{X}_{k+1}^n) \label{bound_o-info}
~,
\end{align}
with $\bs{X}^n_{k} = (X_k,X_{k+1},\dots,X_n)$ and $\bs{X}^k = (X_1,\dots,X_k)$.

Moreover, please note that a tree always has a topological sorting, which guarantees that $\pi(i) < i$ holds. Let us assume, without loss of generality, that the variables are labelled according to a topological sorting. Also, denote as $d(i) = \pi^{-1}(i)$ the set of indices of the descendants of $X_i$ (variables whose parent is $X_i$), which are denoted by $\bs{X}^{d(i)}$.

Note that, by considering Eq.~\eqref{bound_o-info}, $\Omega(\bs{X}^n)\geq 0$ can be proven simply by showing that $I(X_k;\bs{X}^{k-1} ; \bs{X}_{k+1}^n) \geq 0$ for all $1<k<n$. A direct calculation shows that
\begin{align}
I(X_k;\bs{X}^{k-1} &; \bs{X}_{k+1}^n) = I(\bs{X}^{k-1} ; X_k) - I(\bs{X}^{k-1} ; X_k| \bs{X}_{k+1}^n) \nonumber\\
&= I(X_{\pi(k)} ; X_k) - I( X_{\pi(k)}  ; X_k| \bs{X}_{k+1}^n) \nonumber \\
&= I(X_{\pi(k)} ; X_k; \bs{X}_{k+1}^n) \nonumber \\
&= I(X_k ; \bs{X}_{k+1}^n) - I( X_k; \bs{X}_{k+1}^n | X_{\pi(k)} ). \label{eq:ups}
\end{align}
An additional calculation done over the second term from the last expression gives that
\begin{align}
I( X_k; \bs{X}_{k+1}^n | X_{\pi(k)} ) &= I( X_k; \bs{X}^{d(k)} | X_{\pi(k)} ) \nonumber\\
&= I( X_k; \bs{X}^{d(k)}  )  \nonumber\\
&\leq I( X_k; \bs{X}_{k+1}^n  ). \label{eq:yess}
\end{align}
Above, the second equality is a consequence of $X_{\pi(k)} \perp \bs{X}^{d(k)} | X_k$, and the inequality is due to $d(k) \subset \{k+1,\dots,n\}$. By combining Eq.~\eqref{eq:yess} and Eq.~\eqref{eq:ups}, one proves the desired property, which concludes the proof.

\subsubsection*{O-information in Markov chains}
\label{app:markov}

Markov chains are a class of models characterised by the Markov property, for which all the relevant information about $X_{k+1}$ contained in $\bs{X}^{k-1}:=(X_1,\dots, X_{k-1})$ is necessarily contained in $X_k$. Considering a chain $X_1-X_2-\dots-X_n$, this dependency is formalised in the following joint probability distribution:
\begin{equation}
p_{\bs{X}^n}(\bs{x}) = p_{X_1}(x_1) \prod_{i=1}^{n-1}p_{X_{i+1}|X_i}(x_{i+1}|x_i).
\end{equation}
To calculate O-information in this scenario, let us start by calculating $\TC(\bs{X}^n)$. Due to the Markov condition one has that $I(X_i;\bs{X}^{i-1}) = I(X_i;X_{i-1})$. Then, using Eq.~\eqref{bound_TC}, one can show that
\begin{equation}
\TC(\bs{X}^n) = \sum_{i=2}^n I(X_i;X_{i-1})
~.
\end{equation}
This is consistent with the well-known fact that the correlated constraints in Markov systems just include neighbouring variables.

Let us now find an expression for the $\DTC(\bs{X}^n)$. Using Eq.~\eqref{bound_DTC}, and the fact that for Markov chains one has that $I(X_j;\bs{X}^{j-1}|\bs{X}_{j+1}^n) = I(X_j;X_{j-1}|X_{j+1})$, one can show that
\begin{equation}
\DTC(\bs{X}^n) = I(X_n;X_{n-1}) + \sum_{j=2}^{n-1} I(X_j;X_{j-1}|X_{j+1})\label{eqLasdf}
~ .
\end{equation}
Moreover, thanks to the conditions of conditional independence, one can check that
\begin{align}
I(X_j;X_{j-1}|X_{j+1}) =& I(X_j, X_{j+1};X_{j-1}) - I(X_{j+1};X_{j-1}) \nonumber\\
=& I(X_j;X_{j-1}) - I(X_{j+1};X_{j-1}). \label{eqL123}
\end{align}
Using Eq.~\eqref{eqL123} in Eq.~\eqref{eqLasdf}, one finds that
\begin{align}
\DTC(\bs{X}^n) &=  \sum_{j=2}^{n} I(X_j;X_{j-1}) - \sum_{j=2}^{n-1} I(X_{j+1};X_{j-1}) \nonumber \\
&= \TC(\bs{X}^n) - \sum_{j=2}^{n-1} I(X_{j+1};X_{j-1}) . \label{eqLasagfg}
\end{align}
Finally, the desired result follows from re-writing Eq.~\eqref{eqLasagfg} as 
\begin{equation}
\sum_{j=2}^{n-1} I(X_{j+1};X_{j-1})  = \TC(\bs{X}^n) - \DTC(\bs{X}^n) = \Omega(\bs{X}^n).
\end{equation}

\subsubsection*{O-information in Naive Bayes distributions}
\label{app:naivebayes}

Naive Bayes distributions are a type of tree-structured graphical model broadly popular in the machine learning literature because of their tractability, and have been extensively used in domains like genetics, diagnosis, and natural language processing~\cite{Ng2001}. Their generic probability distribution has the form
\begin{equation}
p_{\bs{X}^n}(\bs{x}) = p_{X_n}(x_n) \prod_{i=1}^{n-1} p_{X_i|X_n}(x_i|x_n)~ .
\end{equation}
As for the Markov chain case, let us first find an expression for $\TC(\bs{X}^n)$. For this, let us note that an application of the chain rule shows that
\begin{align}
I(X_i;\bs{X}_{i+1}^n) =& I(X_i;X_n) + I(X_i;\bs{X}_{i+1}^{n-1}| X_n) \nonumber\\
=& I(X_i;X_n). \label{eq:ssaa}
\end{align}
Above, the second equality is a consequence of the conditional independence of $X_i$ and $\bs{X}_{i+1}^{n-1}$ given $X_n$. Using this result on the assembly path decomposition, of the same form as the one given in Eq.~\eqref{bound_TC}, one finds that
\begin{align}
\TC(\bs{X}^n) =& \sum_{i=1}^{n-1} I(X_i;\bs{X}_{i+1}^n) \\
=& \sum_{i=1}^{n-1} I(X_i;X_n) \label{eq1235}
~ .
\end{align}

Let us now find an expression for $\DTC(\bs{X}^n)$. For this, starting from Eq.~\eqref{bound_DTC}, one can show that
\begin{align}
\begin{split}
\DTC(\bs{X}^n) = & I(X_n;\bs{X}^{n-1}) \\
& \quad + \sum_{j=1}^{n-1} I(X_{n-j} ; \bs{X}^{n-j-1} | \bs{X}_{n-j+1}^n) \end{split} \nonumber \\
=& I(X_1;X_n) + \sum_{j=2}^{n-1} I(X_j;X_n|\bs{X}^{j-1}) ~,\label{eqLasdaf}
\end{align}
where the second equality results from the conditional independence of
$X_{n-j}$ and $\bs{X}^{n-j-1}$ given $\bs{X}_{n-j+1}^n$, and an application of the chain rule to $I(X_n;\bs{X}^{n-1})$. Note that
\begin{align}
I(X_j;X_n|\bs{X}^{j-1}) =& I(X_j; \bs{X}^{j-1}, X_n) - I(X_j; \bs{X}^{j-1}) \nonumber \\
=& I(X_j;X_n) - I(X_j; \bs{X}^{j-1})~,
\end{align}
where the second equality is attained by using a derivation similar to
\eqref{eq:ssaa} on the first term. Then, one can rewrite Eq.~\eqref{eqLasdaf} as
\begin{align}
\DTC(\bs{X}^n) =& \sum_{j=1}^{n-1} I(X_j;X_n) - \sum_{k=2}^{n-1} I(X_k; \bs{X}^{k-1}) \\
=& \TC(\bs{X}^n) - \TC(\bs{X}^{n-1}) \label{eq123}
~,
\end{align}
where the second equality is a consequence of Eq.~\eqref{bound_TC}. Finally, the expression for the O-information is attained by noting that, due to Eq.~\eqref{eq123}, one has that
\begin{equation}
\TC(\bs{X}^{n-1})  = \TC(\bs{X}^n) - \DTC(\bs{X}^n) = \Omega(\bs{X}^n).
\end{equation}

\subsubsection*{O-information in stochastic functions}

Outside the class of tree-structured models, graphical models can have negative O-information if variables are allowed to have more than one parent. To show this, let us focus on a simple family of graphical models in which $X_1,\dots, X_{n-1}$ are independent, and $X_n=F(X_1,\dots,X_{n-1})$ with $F(\cdot)$ a (possibly random) function. Then, the joint p.d.f.\ of $\bs{X}^n$ can be written as
\begin{equation}
p_{\bs{X}^n}(\bs{x}^n) = p_{X_n|\bs{X}^{n-1}}(x_n|\bs{x}^{n-1}) \prod_{i=1}^{n-1} p_{X_i}(x_i)\label{eq:indep_methods}
~ .
\end{equation}
To compute $\Omega$, let us first introduce the \textit{conditional dual total
correlation}, given by
\begin{equation}
\DTC(\bs{X}^n|Y) \coloneqq H(\bs{X}^n|Y) - \sum_{j=1}^n H(X_j|\bs{X}^n_{-j},Y)
~ . 
\end{equation}

With this conditional DTC at hand, we can rewrite the O-information (of any system) as
\begin{align} \label{eq:omega_condDTC}
  \Omega(\bs{X}^n) = \TC(\bs{X}^{n-1}) - \DTC(\bs{X}^{n-1}|X_n) ~ .
\end{align}
To show this, note that, using Eqs.~\eqref{bound_TC} and Eq.~\eqref{bound_DTC}, one finds that
\begin{align*}
\TC(\bs{X}^n) &= \TC(\bs{X}^{n-1}) + I(X_n;\bs{X}^{n-1}) ~ , \\
\DTC(\bs{X}^n) &= \DTC(\bs{X}^{n-1}|X_n) + I(X_n;\bs{X}^{n-1}).
\end{align*}
Given that $\TC(\bs{X}^{n-1})=0$ because of independence, the O-information of a system described by Eq.~\eqref{eq:indep_methods} directly follows from Eq.~\eqref{eq:omega_condDTC}:
\begin{equation}
\Omega(\bs{X}^n) = - \DTC(\bs{X}^{n-1}|X_n) \leq 0 ~ .
\end{equation}
This non-positivity of the O-information is governed by interdependencies in $\bs{X}^{n-1}$ which are induced after $X_n$ has been observed. These --- sometimes paradoxical --- interdependencies are known as ``selection bias'' or ``Berkson's paradox''~\cite{berkson1946} in the statistics literature, or as the ``explaining away effect'' in the context of artificial intelligence~\cite{kim1983}.

\section*{Acknowledgements}

F.R. was supported by the Ad Astra Chandaria Foundation, and by the European Union's H2020 research and innovation programme under the Marie Sk\l{}odowska-Curie grant agreement No. 702981.

\bibliography{main}

\providecommand{\noopsort}[1]{}\providecommand{\singleletter}[1]{#1}%
\begin{thebibliography}{85}%
\makeatletter
\providecommand \@ifxundefined [1]{%
 \@ifx{#1\undefined}
}%
\providecommand \@ifnum [1]{%
 \ifnum #1\expandafter \@firstoftwo
 \else \expandafter \@secondoftwo
 \fi
}%
\providecommand \@ifx [1]{%
 \ifx #1\expandafter \@firstoftwo
 \else \expandafter \@secondoftwo
 \fi
}%
\providecommand \natexlab [1]{#1}%
\providecommand \enquote  [1]{``#1''}%
\providecommand \bibnamefont  [1]{#1}%
\providecommand \bibfnamefont [1]{#1}%
\providecommand \citenamefont [1]{#1}%
\providecommand \href@noop [0]{\@secondoftwo}%
\providecommand \href [0]{\begingroup \@sanitize@url \@href}%
\providecommand \@href[1]{\@@startlink{#1}\@@href}%
\providecommand \@@href[1]{\endgroup#1\@@endlink}%
\providecommand \@sanitize@url [0]{\catcode `\\12\catcode `\$12\catcode
  `\&12\catcode `\#12\catcode `\^12\catcode `\_12\catcode `\%12\relax}%
\providecommand \@@startlink[1]{}%
\providecommand \@@endlink[0]{}%
\providecommand \url  [0]{\begingroup\@sanitize@url \@url }%
\providecommand \@url [1]{\endgroup\@href {#1}{\urlprefix }}%
\providecommand \urlprefix  [0]{URL }%
\providecommand \Eprint [0]{\href }%
\providecommand \doibase [0]{http://dx.doi.org/}%
\providecommand \selectlanguage [0]{\@gobble}%
\providecommand \bibinfo  [0]{\@secondoftwo}%
\providecommand \bibfield  [0]{\@secondoftwo}%
\providecommand \translation [1]{[#1]}%
\providecommand \BibitemOpen [0]{}%
\providecommand \bibitemStop [0]{}%
\providecommand \bibitemNoStop [0]{.\EOS\space}%
\providecommand \EOS [0]{\spacefactor3000\relax}%
\providecommand \BibitemShut  [1]{\csname bibitem#1\endcsname}%
\let\auto@bib@innerbib\@empty
\bibitem [{\citenamefont {Waldrop}(1993)}]{waldrop1993complexity}%
  \BibitemOpen
  \bibfield  {author} {\bibinfo {author} {\bibfnamefont {M.~M.}\ \bibnamefont
  {Waldrop}},\ }\href@noop {} {\emph {\bibinfo {title} {Complexity: The
  emerging science at the edge of order and chaos}}}\ (\bibinfo  {publisher}
  {Simon and Schuster},\ \bibinfo {year} {1993})\BibitemShut {NoStop}%
\bibitem [{\citenamefont {Crutchfield}(1994)}]{crutchfield1994calculi}%
  \BibitemOpen
  \bibfield  {author} {\bibinfo {author} {\bibfnamefont {J.~P.}\ \bibnamefont
  {Crutchfield}},\ }\href@noop {} {\bibfield  {journal} {\bibinfo  {journal}
  {Physica D: Nonlinear Phenomena}\ }\textbf {\bibinfo {volume} {75}},\
  \bibinfo {pages} {11} (\bibinfo {year} {1994})}\BibitemShut {NoStop}%
\bibitem [{\citenamefont {Jensen}(2022)}]{jensen2022complexity}%
  \BibitemOpen
  \bibfield  {author} {\bibinfo {author} {\bibfnamefont {H.~J.}\ \bibnamefont
  {Jensen}},\ }\href@noop {} {\emph {\bibinfo {title} {Complexity science: the
  study of emergence}}}\ (\bibinfo  {publisher} {Cambridge University Press},\
  \bibinfo {year} {2022})\BibitemShut {NoStop}%
\bibitem [{\citenamefont {Battiston}\ \emph {et~al.}(2021)\citenamefont
  {Battiston}, \citenamefont {Amico}, \citenamefont {Barrat}, \citenamefont
  {Bianconi}, \citenamefont {Ferraz~de Arruda}, \citenamefont {Franceschiello},
  \citenamefont {Iacopini}, \citenamefont {K{\'e}fi}, \citenamefont {Latora},
  \citenamefont {Moreno} \emph {et~al.}}]{battiston2021physics}%
  \BibitemOpen
  \bibfield  {author} {\bibinfo {author} {\bibfnamefont {F.}~\bibnamefont
  {Battiston}}, \bibinfo {author} {\bibfnamefont {E.}~\bibnamefont {Amico}},
  \bibinfo {author} {\bibfnamefont {A.}~\bibnamefont {Barrat}}, \bibinfo
  {author} {\bibfnamefont {G.}~\bibnamefont {Bianconi}}, \bibinfo {author}
  {\bibfnamefont {G.}~\bibnamefont {Ferraz~de Arruda}}, \bibinfo {author}
  {\bibfnamefont {B.}~\bibnamefont {Franceschiello}}, \bibinfo {author}
  {\bibfnamefont {I.}~\bibnamefont {Iacopini}}, \bibinfo {author}
  {\bibfnamefont {S.}~\bibnamefont {K{\'e}fi}}, \bibinfo {author}
  {\bibfnamefont {V.}~\bibnamefont {Latora}}, \bibinfo {author} {\bibfnamefont
  {Y.}~\bibnamefont {Moreno}},  \emph {et~al.},\ }\href@noop {} {\bibfield
  {journal} {\bibinfo  {journal} {Nature physics}\ }\textbf {\bibinfo {volume}
  {17}},\ \bibinfo {pages} {1093} (\bibinfo {year} {2021})}\BibitemShut
  {NoStop}%
\bibitem [{\citenamefont {Battiston}\ \emph {et~al.}(2026)\citenamefont
  {Battiston}, \citenamefont {Bick}, \citenamefont {Lucas}, \citenamefont
  {Mill{\'a}n}, \citenamefont {Skardal},\ and\ \citenamefont
  {Zhang}}]{battiston2026collective}%
  \BibitemOpen
  \bibfield  {author} {\bibinfo {author} {\bibfnamefont {F.}~\bibnamefont
  {Battiston}}, \bibinfo {author} {\bibfnamefont {C.}~\bibnamefont {Bick}},
  \bibinfo {author} {\bibfnamefont {M.}~\bibnamefont {Lucas}}, \bibinfo
  {author} {\bibfnamefont {A.~P.}\ \bibnamefont {Mill{\'a}n}}, \bibinfo
  {author} {\bibfnamefont {P.~S.}\ \bibnamefont {Skardal}}, \ and\ \bibinfo
  {author} {\bibfnamefont {Y.}~\bibnamefont {Zhang}},\ }\href@noop {}
  {\bibfield  {journal} {\bibinfo  {journal} {Nature Reviews Physics}\ }\textbf
  {\bibinfo {volume} {8}},\ \bibinfo {pages} {146} (\bibinfo {year}
  {2026})}\BibitemShut {NoStop}%
\bibitem [{\citenamefont {Rajpal}\ \emph {et~al.}(2023)\citenamefont {Rajpal},
  \citenamefont {von Stengel}, \citenamefont {Mediano}, \citenamefont {Rosas},
  \citenamefont {Viegas}, \citenamefont {Marquet},\ and\ \citenamefont
  {Jensen}}]{rajpal2023quantifying}%
  \BibitemOpen
  \bibfield  {author} {\bibinfo {author} {\bibfnamefont {H.}~\bibnamefont
  {Rajpal}}, \bibinfo {author} {\bibfnamefont {C.}~\bibnamefont {von Stengel}},
  \bibinfo {author} {\bibfnamefont {P.~A.}\ \bibnamefont {Mediano}}, \bibinfo
  {author} {\bibfnamefont {F.~E.}\ \bibnamefont {Rosas}}, \bibinfo {author}
  {\bibfnamefont {E.}~\bibnamefont {Viegas}}, \bibinfo {author} {\bibfnamefont
  {P.~A.}\ \bibnamefont {Marquet}}, \ and\ \bibinfo {author} {\bibfnamefont
  {H.~J.}\ \bibnamefont {Jensen}},\ }\href@noop {} {\bibfield  {journal}
  {\bibinfo  {journal} {arXiv:2310.20386}\ } (\bibinfo {year}
  {2023})}\BibitemShut {NoStop}%
\bibitem [{\citenamefont {Cang}\ and\ \citenamefont
  {Nie}(2020)}]{cang2020inferring}%
  \BibitemOpen
  \bibfield  {author} {\bibinfo {author} {\bibfnamefont {Z.}~\bibnamefont
  {Cang}}\ and\ \bibinfo {author} {\bibfnamefont {Q.}~\bibnamefont {Nie}},\
  }\href@noop {} {\bibfield  {journal} {\bibinfo  {journal} {Nature
  communications}\ }\textbf {\bibinfo {volume} {11}},\ \bibinfo {pages} {2084}
  (\bibinfo {year} {2020})}\BibitemShut {NoStop}%
\bibitem [{\citenamefont {Battiston}\ \emph {et~al.}(2025)\citenamefont
  {Battiston}, \citenamefont {Capraro}, \citenamefont {Karimi}, \citenamefont
  {Lehmann}, \citenamefont {Migliano}, \citenamefont {Sadekar}, \citenamefont
  {S{\'a}nchez},\ and\ \citenamefont {Perc}}]{battiston2025higher}%
  \BibitemOpen
  \bibfield  {author} {\bibinfo {author} {\bibfnamefont {F.}~\bibnamefont
  {Battiston}}, \bibinfo {author} {\bibfnamefont {V.}~\bibnamefont {Capraro}},
  \bibinfo {author} {\bibfnamefont {F.}~\bibnamefont {Karimi}}, \bibinfo
  {author} {\bibfnamefont {S.}~\bibnamefont {Lehmann}}, \bibinfo {author}
  {\bibfnamefont {A.~B.}\ \bibnamefont {Migliano}}, \bibinfo {author}
  {\bibfnamefont {O.}~\bibnamefont {Sadekar}}, \bibinfo {author} {\bibfnamefont
  {A.}~\bibnamefont {S{\'a}nchez}}, \ and\ \bibinfo {author} {\bibfnamefont
  {M.}~\bibnamefont {Perc}},\ }\href@noop {} {\bibfield  {journal} {\bibinfo
  {journal} {Nature Human Behaviour}\ ,\ \bibinfo {pages} {1}} (\bibinfo {year}
  {2025})}\BibitemShut {NoStop}%
\bibitem [{\citenamefont {Marinazzo}\ \emph {et~al.}(2024)\citenamefont
  {Marinazzo}, \citenamefont {Van~Roozendaal}, \citenamefont {Rosas},
  \citenamefont {Stella}, \citenamefont {Comolatti}, \citenamefont {Colenbier},
  \citenamefont {Stramaglia},\ and\ \citenamefont
  {Rosseel}}]{marinazzo2024information}%
  \BibitemOpen
  \bibfield  {author} {\bibinfo {author} {\bibfnamefont {D.}~\bibnamefont
  {Marinazzo}}, \bibinfo {author} {\bibfnamefont {J.}~\bibnamefont
  {Van~Roozendaal}}, \bibinfo {author} {\bibfnamefont {F.~E.}\ \bibnamefont
  {Rosas}}, \bibinfo {author} {\bibfnamefont {M.}~\bibnamefont {Stella}},
  \bibinfo {author} {\bibfnamefont {R.}~\bibnamefont {Comolatti}}, \bibinfo
  {author} {\bibfnamefont {N.}~\bibnamefont {Colenbier}}, \bibinfo {author}
  {\bibfnamefont {S.}~\bibnamefont {Stramaglia}}, \ and\ \bibinfo {author}
  {\bibfnamefont {Y.}~\bibnamefont {Rosseel}},\ }\href@noop {} {\bibfield
  {journal} {\bibinfo  {journal} {Behavior Research Methods}\ }\textbf
  {\bibinfo {volume} {56}},\ \bibinfo {pages} {8057} (\bibinfo {year}
  {2024})}\BibitemShut {NoStop}%
\bibitem [{\citenamefont {Varley}\ and\ \citenamefont
  {Kaminski}(2022)}]{varley2022untangling}%
  \BibitemOpen
  \bibfield  {author} {\bibinfo {author} {\bibfnamefont {T.~F.}\ \bibnamefont
  {Varley}}\ and\ \bibinfo {author} {\bibfnamefont {P.}~\bibnamefont
  {Kaminski}},\ }\href@noop {} {\bibfield  {journal} {\bibinfo  {journal}
  {Entropy}\ }\textbf {\bibinfo {volume} {24}},\ \bibinfo {pages} {1387}
  (\bibinfo {year} {2022})}\BibitemShut {NoStop}%
\bibitem [{\citenamefont {Rosas}\ \emph {et~al.}(2018)\citenamefont {Rosas},
  \citenamefont {Mediano}, \citenamefont {Ugarte},\ and\ \citenamefont
  {Jensen}}]{Rosas2018}%
  \BibitemOpen
  \bibfield  {author} {\bibinfo {author} {\bibfnamefont {F.~E.}\ \bibnamefont
  {Rosas}}, \bibinfo {author} {\bibfnamefont {P.}~\bibnamefont {Mediano}},
  \bibinfo {author} {\bibfnamefont {M.}~\bibnamefont {Ugarte}}, \ and\ \bibinfo
  {author} {\bibfnamefont {H.}~\bibnamefont {Jensen}},\ }\href@noop {}
  {\bibfield  {journal} {\bibinfo  {journal} {Entropy}\ }\textbf {\bibinfo
  {volume} {20}},\ \bibinfo {pages} {793} (\bibinfo {year} {2018})}\BibitemShut
  {NoStop}%
\bibitem [{\citenamefont {Tax}\ \emph {et~al.}(2017)\citenamefont {Tax},
  \citenamefont {Mediano},\ and\ \citenamefont {Shanahan}}]{tax2017partial}%
  \BibitemOpen
  \bibfield  {author} {\bibinfo {author} {\bibfnamefont {T.~M.}\ \bibnamefont
  {Tax}}, \bibinfo {author} {\bibfnamefont {P.~A.}\ \bibnamefont {Mediano}}, \
  and\ \bibinfo {author} {\bibfnamefont {M.}~\bibnamefont {Shanahan}},\
  }\href@noop {} {\bibfield  {journal} {\bibinfo  {journal} {Entropy}\ }\textbf
  {\bibinfo {volume} {19}},\ \bibinfo {pages} {474} (\bibinfo {year}
  {2017})}\BibitemShut {NoStop}%
\bibitem [{\citenamefont {Proca}\ \emph {et~al.}(2024)\citenamefont {Proca},
  \citenamefont {Rosas}, \citenamefont {Luppi}, \citenamefont {Bor},
  \citenamefont {Crosby},\ and\ \citenamefont
  {Mediano}}]{proca2024synergistic}%
  \BibitemOpen
  \bibfield  {author} {\bibinfo {author} {\bibfnamefont {A.~M.}\ \bibnamefont
  {Proca}}, \bibinfo {author} {\bibfnamefont {F.~E.}\ \bibnamefont {Rosas}},
  \bibinfo {author} {\bibfnamefont {A.~I.}\ \bibnamefont {Luppi}}, \bibinfo
  {author} {\bibfnamefont {D.}~\bibnamefont {Bor}}, \bibinfo {author}
  {\bibfnamefont {M.}~\bibnamefont {Crosby}}, \ and\ \bibinfo {author}
  {\bibfnamefont {P.~A.}\ \bibnamefont {Mediano}},\ }\href@noop {} {\bibfield
  {journal} {\bibinfo  {journal} {PLOS Computational Biology}\ }\textbf
  {\bibinfo {volume} {20}},\ \bibinfo {pages} {e1012178} (\bibinfo {year}
  {2024})}\BibitemShut {NoStop}%
\bibitem [{\citenamefont {Kaplanis}\ \emph {et~al.}(2023)\citenamefont
  {Kaplanis}, \citenamefont {Mediano},\ and\ \citenamefont
  {Rosas}}]{kaplanis2023learning}%
  \BibitemOpen
  \bibfield  {author} {\bibinfo {author} {\bibfnamefont {C.}~\bibnamefont
  {Kaplanis}}, \bibinfo {author} {\bibfnamefont {P.}~\bibnamefont {Mediano}}, \
  and\ \bibinfo {author} {\bibfnamefont {F.}~\bibnamefont {Rosas}},\ }in\
  \href@noop {} {\emph {\bibinfo {booktitle} {NeurIPS 2023 workshop:
  Information-Theoretic Principles in Cognitive Systems}}}\ (\bibinfo {year}
  {2023})\BibitemShut {NoStop}%
\bibitem [{\citenamefont {Gutknecht}\ \emph {et~al.}(2025)\citenamefont
  {Gutknecht}, \citenamefont {Rosas}, \citenamefont {Ehrlich}, \citenamefont
  {Makkeh}, \citenamefont {Mediano},\ and\ \citenamefont
  {Wibral}}]{gutknecht2025shannon}%
  \BibitemOpen
  \bibfield  {author} {\bibinfo {author} {\bibfnamefont {A.~J.}\ \bibnamefont
  {Gutknecht}}, \bibinfo {author} {\bibfnamefont {F.~E.}\ \bibnamefont
  {Rosas}}, \bibinfo {author} {\bibfnamefont {D.~A.}\ \bibnamefont {Ehrlich}},
  \bibinfo {author} {\bibfnamefont {A.}~\bibnamefont {Makkeh}}, \bibinfo
  {author} {\bibfnamefont {P.~A.~M.}\ \bibnamefont {Mediano}}, \ and\ \bibinfo
  {author} {\bibfnamefont {M.}~\bibnamefont {Wibral}},\ }\href@noop {}
  {\bibfield  {journal} {\bibinfo  {journal} {arXiv preprint arXiv:2504.15779}\
  } (\bibinfo {year} {2025})}\BibitemShut {NoStop}%
\bibitem [{\citenamefont {Gatica}\ \emph {et~al.}(2021)\citenamefont {Gatica},
  \citenamefont {Cofr{\'e}}, \citenamefont {Mediano}, \citenamefont {Rosas},
  \citenamefont {Orio}, \citenamefont {Diez}, \citenamefont {Swinnen},\ and\
  \citenamefont {Cortes}}]{gatica2021high}%
  \BibitemOpen
  \bibfield  {author} {\bibinfo {author} {\bibfnamefont {M.}~\bibnamefont
  {Gatica}}, \bibinfo {author} {\bibfnamefont {R.}~\bibnamefont {Cofr{\'e}}},
  \bibinfo {author} {\bibfnamefont {P.~A.}\ \bibnamefont {Mediano}}, \bibinfo
  {author} {\bibfnamefont {F.~E.}\ \bibnamefont {Rosas}}, \bibinfo {author}
  {\bibfnamefont {P.}~\bibnamefont {Orio}}, \bibinfo {author} {\bibfnamefont
  {I.}~\bibnamefont {Diez}}, \bibinfo {author} {\bibfnamefont {S.~P.}\
  \bibnamefont {Swinnen}}, \ and\ \bibinfo {author} {\bibfnamefont {J.~M.}\
  \bibnamefont {Cortes}},\ }\href@noop {} {\bibfield  {journal} {\bibinfo
  {journal} {Brain connectivity}\ }\textbf {\bibinfo {volume} {11}},\ \bibinfo
  {pages} {734} (\bibinfo {year} {2021})}\BibitemShut {NoStop}%
\bibitem [{\citenamefont {Luppi}\ \emph {et~al.}(2022)\citenamefont {Luppi},
  \citenamefont {Mediano}, \citenamefont {Rosas}, \citenamefont {Holland},
  \citenamefont {Fryer}, \citenamefont {O’Brien}, \citenamefont {Rowe},
  \citenamefont {Menon}, \citenamefont {Bor},\ and\ \citenamefont
  {Stamatakis}}]{luppi2022synergistic}%
  \BibitemOpen
  \bibfield  {author} {\bibinfo {author} {\bibfnamefont {A.~I.}\ \bibnamefont
  {Luppi}}, \bibinfo {author} {\bibfnamefont {P.~A.}\ \bibnamefont {Mediano}},
  \bibinfo {author} {\bibfnamefont {F.~E.}\ \bibnamefont {Rosas}}, \bibinfo
  {author} {\bibfnamefont {N.}~\bibnamefont {Holland}}, \bibinfo {author}
  {\bibfnamefont {T.~D.}\ \bibnamefont {Fryer}}, \bibinfo {author}
  {\bibfnamefont {J.~T.}\ \bibnamefont {O’Brien}}, \bibinfo {author}
  {\bibfnamefont {J.~B.}\ \bibnamefont {Rowe}}, \bibinfo {author}
  {\bibfnamefont {D.~K.}\ \bibnamefont {Menon}}, \bibinfo {author}
  {\bibfnamefont {D.}~\bibnamefont {Bor}}, \ and\ \bibinfo {author}
  {\bibfnamefont {E.~A.}\ \bibnamefont {Stamatakis}},\ }\href@noop {}
  {\bibfield  {journal} {\bibinfo  {journal} {Nature Neuroscience}\ ,\ \bibinfo
  {pages} {1}} (\bibinfo {year} {2022})}\BibitemShut {NoStop}%
\bibitem [{\citenamefont {Herzog}\ \emph {et~al.}(2022)\citenamefont {Herzog},
  \citenamefont {Rosas}, \citenamefont {Whelan}, \citenamefont {Fittipaldi},
  \citenamefont {Santamaria-Garcia}, \citenamefont {Cruzat}, \citenamefont
  {Birba}, \citenamefont {Moguilner}, \citenamefont {Tagliazucchi},
  \citenamefont {Prado} \emph {et~al.}}]{herzog2022genuine}%
  \BibitemOpen
  \bibfield  {author} {\bibinfo {author} {\bibfnamefont {R.}~\bibnamefont
  {Herzog}}, \bibinfo {author} {\bibfnamefont {F.~E.}\ \bibnamefont {Rosas}},
  \bibinfo {author} {\bibfnamefont {R.}~\bibnamefont {Whelan}}, \bibinfo
  {author} {\bibfnamefont {S.}~\bibnamefont {Fittipaldi}}, \bibinfo {author}
  {\bibfnamefont {H.}~\bibnamefont {Santamaria-Garcia}}, \bibinfo {author}
  {\bibfnamefont {J.}~\bibnamefont {Cruzat}}, \bibinfo {author} {\bibfnamefont
  {A.}~\bibnamefont {Birba}}, \bibinfo {author} {\bibfnamefont
  {S.}~\bibnamefont {Moguilner}}, \bibinfo {author} {\bibfnamefont
  {E.}~\bibnamefont {Tagliazucchi}}, \bibinfo {author} {\bibfnamefont
  {P.}~\bibnamefont {Prado}},  \emph {et~al.},\ }\href@noop {} {\bibfield
  {journal} {\bibinfo  {journal} {Neurobiology of Disease}\ }\textbf {\bibinfo
  {volume} {175}},\ \bibinfo {pages} {105918} (\bibinfo {year}
  {2022})}\BibitemShut {NoStop}%
\bibitem [{\citenamefont {Varley}\ \emph
  {et~al.}(2023{\natexlab{a}})\citenamefont {Varley}, \citenamefont {Sporns},
  \citenamefont {Schaffelhofer}, \citenamefont {Scherberger},\ and\
  \citenamefont {Dann}}]{varley2023information}%
  \BibitemOpen
  \bibfield  {author} {\bibinfo {author} {\bibfnamefont {T.~F.}\ \bibnamefont
  {Varley}}, \bibinfo {author} {\bibfnamefont {O.}~\bibnamefont {Sporns}},
  \bibinfo {author} {\bibfnamefont {S.}~\bibnamefont {Schaffelhofer}}, \bibinfo
  {author} {\bibfnamefont {H.}~\bibnamefont {Scherberger}}, \ and\ \bibinfo
  {author} {\bibfnamefont {B.}~\bibnamefont {Dann}},\ }\href@noop {} {\bibfield
   {journal} {\bibinfo  {journal} {Proceedings of the National Academy of
  Sciences}\ }\textbf {\bibinfo {volume} {120}},\ \bibinfo {pages}
  {e2207677120} (\bibinfo {year} {2023}{\natexlab{a}})}\BibitemShut {NoStop}%
\bibitem [{\citenamefont {Varley}\ \emph
  {et~al.}(2023{\natexlab{b}})\citenamefont {Varley}, \citenamefont {Pope},
  \citenamefont {Faskowitz},\ and\ \citenamefont
  {Sporns}}]{varley2023multivariate}%
  \BibitemOpen
  \bibfield  {author} {\bibinfo {author} {\bibfnamefont {T.~F.}\ \bibnamefont
  {Varley}}, \bibinfo {author} {\bibfnamefont {M.}~\bibnamefont {Pope}},
  \bibinfo {author} {\bibfnamefont {J.}~\bibnamefont {Faskowitz}}, \ and\
  \bibinfo {author} {\bibfnamefont {O.}~\bibnamefont {Sporns}},\ }\href@noop {}
  {\bibfield  {journal} {\bibinfo  {journal} {Communications biology}\ }\textbf
  {\bibinfo {volume} {6}},\ \bibinfo {pages} {451} (\bibinfo {year}
  {2023}{\natexlab{b}})}\BibitemShut {NoStop}%
\bibitem [{\citenamefont {Rosas}\ \emph {et~al.}(2022)\citenamefont {Rosas},
  \citenamefont {Mediano}, \citenamefont {Luppi}, \citenamefont {Varley},
  \citenamefont {Lizier}, \citenamefont {Stramaglia}, \citenamefont {Jensen},\
  and\ \citenamefont {Marinazzo}}]{rosas2022disentangling}%
  \BibitemOpen
  \bibfield  {author} {\bibinfo {author} {\bibfnamefont {F.~E.}\ \bibnamefont
  {Rosas}}, \bibinfo {author} {\bibfnamefont {P.~A.}\ \bibnamefont {Mediano}},
  \bibinfo {author} {\bibfnamefont {A.~I.}\ \bibnamefont {Luppi}}, \bibinfo
  {author} {\bibfnamefont {T.~F.}\ \bibnamefont {Varley}}, \bibinfo {author}
  {\bibfnamefont {J.~T.}\ \bibnamefont {Lizier}}, \bibinfo {author}
  {\bibfnamefont {S.}~\bibnamefont {Stramaglia}}, \bibinfo {author}
  {\bibfnamefont {H.~J.}\ \bibnamefont {Jensen}}, \ and\ \bibinfo {author}
  {\bibfnamefont {D.}~\bibnamefont {Marinazzo}},\ }\href@noop {} {\bibfield
  {journal} {\bibinfo  {journal} {Nature Physics}\ }\textbf {\bibinfo {volume}
  {18}},\ \bibinfo {pages} {476} (\bibinfo {year} {2022})}\BibitemShut
  {NoStop}%
\bibitem [{\citenamefont {Varley}\ \emph
  {et~al.}(2023{\natexlab{c}})\citenamefont {Varley}, \citenamefont {Pope},
  \citenamefont {Grazia}, \citenamefont {Joshua},\ and\ \citenamefont
  {Sporns}}]{varley2023partial}%
  \BibitemOpen
  \bibfield  {author} {\bibinfo {author} {\bibfnamefont {T.~F.}\ \bibnamefont
  {Varley}}, \bibinfo {author} {\bibfnamefont {M.}~\bibnamefont {Pope}},
  \bibinfo {author} {\bibfnamefont {M.}~\bibnamefont {Grazia}}, \bibinfo
  {author} {\bibnamefont {Joshua}}, \ and\ \bibinfo {author} {\bibfnamefont
  {O.}~\bibnamefont {Sporns}},\ }\href@noop {} {\bibfield  {journal} {\bibinfo
  {journal} {Proceedings of the National Academy of Sciences}\ }\textbf
  {\bibinfo {volume} {120}},\ \bibinfo {pages} {e2300888120} (\bibinfo {year}
  {2023}{\natexlab{c}})}\BibitemShut {NoStop}%
\bibitem [{\citenamefont {Williams}\ and\ \citenamefont
  {Beer}(2010)}]{williams2010nonnegative}%
  \BibitemOpen
  \bibfield  {author} {\bibinfo {author} {\bibfnamefont {P.~L.}\ \bibnamefont
  {Williams}}\ and\ \bibinfo {author} {\bibfnamefont {R.~D.}\ \bibnamefont
  {Beer}},\ }\href@noop {} {\bibfield  {journal} {\bibinfo  {journal} {arXiv
  preprint arXiv:1004.2515}\ } (\bibinfo {year} {2010})}\BibitemShut {NoStop}%
\bibitem [{\citenamefont {Mediano}\ \emph {et~al.}(2025)\citenamefont
  {Mediano}, \citenamefont {Rosas}, \citenamefont {Luppi}, \citenamefont
  {Carhart-Harris}, \citenamefont {Bor}, \citenamefont {Seth},\ and\
  \citenamefont {Barrett}}]{mediano2025toward}%
  \BibitemOpen
  \bibfield  {author} {\bibinfo {author} {\bibfnamefont {P.~A.}\ \bibnamefont
  {Mediano}}, \bibinfo {author} {\bibfnamefont {F.~E.}\ \bibnamefont {Rosas}},
  \bibinfo {author} {\bibfnamefont {A.~I.}\ \bibnamefont {Luppi}}, \bibinfo
  {author} {\bibfnamefont {R.~L.}\ \bibnamefont {Carhart-Harris}}, \bibinfo
  {author} {\bibfnamefont {D.}~\bibnamefont {Bor}}, \bibinfo {author}
  {\bibfnamefont {A.~K.}\ \bibnamefont {Seth}}, \ and\ \bibinfo {author}
  {\bibfnamefont {A.~B.}\ \bibnamefont {Barrett}},\ }\href@noop {} {\bibfield
  {journal} {\bibinfo  {journal} {Proceedings of the National Academy of
  Sciences}\ }\textbf {\bibinfo {volume} {122}},\ \bibinfo {pages}
  {e2423297122} (\bibinfo {year} {2025})}\BibitemShut {NoStop}%
\bibitem [{\citenamefont {Ince}(2017)}]{ince2017partial}%
  \BibitemOpen
  \bibfield  {author} {\bibinfo {author} {\bibfnamefont {R.~A.}\ \bibnamefont
  {Ince}},\ }\href@noop {} {\bibfield  {journal} {\bibinfo  {journal} {arXiv
  preprint arXiv:1702.01591}\ } (\bibinfo {year} {2017})}\BibitemShut {NoStop}%
\bibitem [{\citenamefont {Rosas}\ \emph {et~al.}(2025)\citenamefont {Rosas},
  \citenamefont {Gutknecht}, \citenamefont {Mediano},\ and\ \citenamefont
  {Gastpar}}]{rosas2025characterising}%
  \BibitemOpen
  \bibfield  {author} {\bibinfo {author} {\bibfnamefont {F.~E.}\ \bibnamefont
  {Rosas}}, \bibinfo {author} {\bibfnamefont {A.~J.}\ \bibnamefont
  {Gutknecht}}, \bibinfo {author} {\bibfnamefont {P.~A.}\ \bibnamefont
  {Mediano}}, \ and\ \bibinfo {author} {\bibfnamefont {M.}~\bibnamefont
  {Gastpar}},\ }\href@noop {} {\bibfield  {journal} {\bibinfo  {journal}
  {Communications Physics}\ }\textbf {\bibinfo {volume} {8}},\ \bibinfo {pages}
  {347} (\bibinfo {year} {2025})}\BibitemShut {NoStop}%
\bibitem [{\citenamefont {Liardi}\ \emph {et~al.}(2026)\citenamefont {Liardi},
  \citenamefont {Down}, \citenamefont {Blackburne}, \citenamefont {Neri},\ and\
  \citenamefont {Mediano}}]{liardi2026mathematical}%
  \BibitemOpen
  \bibfield  {author} {\bibinfo {author} {\bibfnamefont {A.}~\bibnamefont
  {Liardi}}, \bibinfo {author} {\bibfnamefont {K.~J.}\ \bibnamefont {Down}},
  \bibinfo {author} {\bibfnamefont {G.}~\bibnamefont {Blackburne}}, \bibinfo
  {author} {\bibfnamefont {M.}~\bibnamefont {Neri}}, \ and\ \bibinfo {author}
  {\bibfnamefont {P.~A.}\ \bibnamefont {Mediano}},\ }\href@noop {} {\bibfield
  {journal} {\bibinfo  {journal} {arXiv preprint arXiv:2603.06678}\ } (\bibinfo
  {year} {2026})}\BibitemShut {NoStop}%
\bibitem [{\citenamefont {Luppi}\ \emph {et~al.}(2024)\citenamefont {Luppi},
  \citenamefont {Mediano}, \citenamefont {Rosas}, \citenamefont {Allanson},
  \citenamefont {Pickard}, \citenamefont {Carhart-Harris}, \citenamefont
  {Williams}, \citenamefont {Craig}, \citenamefont {Finoia}, \citenamefont
  {Owen} \emph {et~al.}}]{luppi2024synergistic}%
  \BibitemOpen
  \bibfield  {author} {\bibinfo {author} {\bibfnamefont {A.~I.}\ \bibnamefont
  {Luppi}}, \bibinfo {author} {\bibfnamefont {P.~A.}\ \bibnamefont {Mediano}},
  \bibinfo {author} {\bibfnamefont {F.~E.}\ \bibnamefont {Rosas}}, \bibinfo
  {author} {\bibfnamefont {J.}~\bibnamefont {Allanson}}, \bibinfo {author}
  {\bibfnamefont {J.}~\bibnamefont {Pickard}}, \bibinfo {author} {\bibfnamefont
  {R.~L.}\ \bibnamefont {Carhart-Harris}}, \bibinfo {author} {\bibfnamefont
  {G.~B.}\ \bibnamefont {Williams}}, \bibinfo {author} {\bibfnamefont {M.~M.}\
  \bibnamefont {Craig}}, \bibinfo {author} {\bibfnamefont {P.}~\bibnamefont
  {Finoia}}, \bibinfo {author} {\bibfnamefont {A.~M.}\ \bibnamefont {Owen}},
  \emph {et~al.},\ }\href@noop {} {\bibfield  {journal} {\bibinfo  {journal}
  {Elife}\ }\textbf {\bibinfo {volume} {12}},\ \bibinfo {pages} {RP88173}
  (\bibinfo {year} {2024})}\BibitemShut {NoStop}%
\bibitem [{\citenamefont {Liardi}\ \emph
  {et~al.}(2025{\natexlab{a}})\citenamefont {Liardi}, \citenamefont {Rosas},
  \citenamefont {Carhart-Harris}, \citenamefont {Blackburne}, \citenamefont
  {Bor},\ and\ \citenamefont {Mediano}}]{liardi2025null}%
  \BibitemOpen
  \bibfield  {author} {\bibinfo {author} {\bibfnamefont {A.}~\bibnamefont
  {Liardi}}, \bibinfo {author} {\bibfnamefont {F.~E.}\ \bibnamefont {Rosas}},
  \bibinfo {author} {\bibfnamefont {R.~L.}\ \bibnamefont {Carhart-Harris}},
  \bibinfo {author} {\bibfnamefont {G.}~\bibnamefont {Blackburne}}, \bibinfo
  {author} {\bibfnamefont {D.}~\bibnamefont {Bor}}, \ and\ \bibinfo {author}
  {\bibfnamefont {P.~A.}\ \bibnamefont {Mediano}},\ }\href@noop {} {\bibfield
  {journal} {\bibinfo  {journal} {PLOS Computational Biology}\ }\textbf
  {\bibinfo {volume} {21}},\ \bibinfo {pages} {e1013629} (\bibinfo {year}
  {2025}{\natexlab{a}})}\BibitemShut {NoStop}%
\bibitem [{\citenamefont {Goodwell}\ and\ \citenamefont
  {Kumar}(2017)}]{goodwell2017temporal}%
  \BibitemOpen
  \bibfield  {author} {\bibinfo {author} {\bibfnamefont {A.~E.}\ \bibnamefont
  {Goodwell}}\ and\ \bibinfo {author} {\bibfnamefont {P.}~\bibnamefont
  {Kumar}},\ }\href@noop {} {\bibfield  {journal} {\bibinfo  {journal} {Water
  Resources Research}\ }\textbf {\bibinfo {volume} {53}},\ \bibinfo {pages}
  {5920} (\bibinfo {year} {2017})}\BibitemShut {NoStop}%
\bibitem [{\citenamefont {Mediano}\ \emph {et~al.}(2022)\citenamefont
  {Mediano}, \citenamefont {Rosas}, \citenamefont {Farah}, \citenamefont
  {Shanahan}, \citenamefont {Bor},\ and\ \citenamefont
  {Barrett}}]{mediano2022integrated}%
  \BibitemOpen
  \bibfield  {author} {\bibinfo {author} {\bibfnamefont {P.~A.}\ \bibnamefont
  {Mediano}}, \bibinfo {author} {\bibfnamefont {F.~E.}\ \bibnamefont {Rosas}},
  \bibinfo {author} {\bibfnamefont {J.~C.}\ \bibnamefont {Farah}}, \bibinfo
  {author} {\bibfnamefont {M.}~\bibnamefont {Shanahan}}, \bibinfo {author}
  {\bibfnamefont {D.}~\bibnamefont {Bor}}, \ and\ \bibinfo {author}
  {\bibfnamefont {A.~B.}\ \bibnamefont {Barrett}},\ }\href@noop {} {\bibfield
  {journal} {\bibinfo  {journal} {Chaos: An Interdisciplinary Journal of
  Nonlinear Science}\ }\textbf {\bibinfo {volume} {32}},\ \bibinfo {pages}
  {013115} (\bibinfo {year} {2022})}\BibitemShut {NoStop}%
\bibitem [{\citenamefont {Marinazzo}\ \emph {et~al.}(2019)\citenamefont
  {Marinazzo}, \citenamefont {Angelini}, \citenamefont {Pellicoro},\ and\
  \citenamefont {Stramaglia}}]{marinazzo2019synergy}%
  \BibitemOpen
  \bibfield  {author} {\bibinfo {author} {\bibfnamefont {D.}~\bibnamefont
  {Marinazzo}}, \bibinfo {author} {\bibfnamefont {L.}~\bibnamefont {Angelini}},
  \bibinfo {author} {\bibfnamefont {M.}~\bibnamefont {Pellicoro}}, \ and\
  \bibinfo {author} {\bibfnamefont {S.}~\bibnamefont {Stramaglia}},\
  }\href@noop {} {\bibfield  {journal} {\bibinfo  {journal} {Physical Review
  E}\ }\textbf {\bibinfo {volume} {99}},\ \bibinfo {pages} {040101} (\bibinfo
  {year} {2019})}\BibitemShut {NoStop}%
\bibitem [{\citenamefont {Caprioglio}\ \emph {et~al.}(2026)\citenamefont
  {Caprioglio}, \citenamefont {Mediano},\ and\ \citenamefont
  {Berthouze}}]{caprioglio2025synergistic}%
  \BibitemOpen
  \bibfield  {author} {\bibinfo {author} {\bibfnamefont {E.}~\bibnamefont
  {Caprioglio}}, \bibinfo {author} {\bibfnamefont {P.~A.}\ \bibnamefont
  {Mediano}}, \ and\ \bibinfo {author} {\bibfnamefont {L.}~\bibnamefont
  {Berthouze}},\ }\href@noop {} {\bibfield  {journal} {\bibinfo  {journal}
  {Physical Review Letters}\ }\textbf {\bibinfo {volume} {136}},\ \bibinfo
  {pages} {127401} (\bibinfo {year} {2026})}\BibitemShut {NoStop}%
\bibitem [{\citenamefont {Rajpal}\ and\ \citenamefont
  {Guerrero}(2025)}]{rajpal2025synergistic}%
  \BibitemOpen
  \bibfield  {author} {\bibinfo {author} {\bibfnamefont {H.}~\bibnamefont
  {Rajpal}}\ and\ \bibinfo {author} {\bibfnamefont {O.}~\bibnamefont
  {Guerrero}},\ }\href@noop {} {\bibfield  {journal} {\bibinfo  {journal} {PNAS
  nexus}\ }\textbf {\bibinfo {volume} {4}},\ \bibinfo {pages} {pgaf102}
  (\bibinfo {year} {2025})}\BibitemShut {NoStop}%
\bibitem [{\citenamefont {Sas}\ \emph {et~al.}(2025)\citenamefont {Sas},
  \citenamefont {Rosas}, \citenamefont {Rajpal}, \citenamefont {Bor},
  \citenamefont {Jensen},\ and\ \citenamefont {Mediano}}]{sas2025improved}%
  \BibitemOpen
  \bibfield  {author} {\bibinfo {author} {\bibfnamefont {M.~I.}\ \bibnamefont
  {Sas}}, \bibinfo {author} {\bibfnamefont {F.~E.}\ \bibnamefont {Rosas}},
  \bibinfo {author} {\bibfnamefont {H.}~\bibnamefont {Rajpal}}, \bibinfo
  {author} {\bibfnamefont {D.}~\bibnamefont {Bor}}, \bibinfo {author}
  {\bibfnamefont {H.~J.}\ \bibnamefont {Jensen}}, \ and\ \bibinfo {author}
  {\bibfnamefont {P.~A.}\ \bibnamefont {Mediano}},\ }\href@noop {} {\bibfield
  {journal} {\bibinfo  {journal} {arXiv preprint arXiv:2601.00013}\ } (\bibinfo
  {year} {2025})}\BibitemShut {NoStop}%
\bibitem [{\citenamefont {Varley}\ \emph {et~al.}(2025)\citenamefont {Varley},
  \citenamefont {Mediano}, \citenamefont {Patania},\ and\ \citenamefont
  {Bongard}}]{varley2025topology}%
  \BibitemOpen
  \bibfield  {author} {\bibinfo {author} {\bibfnamefont {T.~F.}\ \bibnamefont
  {Varley}}, \bibinfo {author} {\bibfnamefont {P.~A.}\ \bibnamefont {Mediano}},
  \bibinfo {author} {\bibfnamefont {A.}~\bibnamefont {Patania}}, \ and\
  \bibinfo {author} {\bibfnamefont {J.}~\bibnamefont {Bongard}},\ }\href@noop
  {} {\bibfield  {journal} {\bibinfo  {journal} {PLOS Computational Biology}\
  }\textbf {\bibinfo {volume} {21}},\ \bibinfo {pages} {e1013649} (\bibinfo
  {year} {2025})}\BibitemShut {NoStop}%
\bibitem [{\citenamefont {Scagliarini}\ \emph {et~al.}(2022)\citenamefont
  {Scagliarini}, \citenamefont {Marinazzo}, \citenamefont {Guo}, \citenamefont
  {Stramaglia},\ and\ \citenamefont {Rosas}}]{scagliarini2022quantifying}%
  \BibitemOpen
  \bibfield  {author} {\bibinfo {author} {\bibfnamefont {T.}~\bibnamefont
  {Scagliarini}}, \bibinfo {author} {\bibfnamefont {D.}~\bibnamefont
  {Marinazzo}}, \bibinfo {author} {\bibfnamefont {Y.}~\bibnamefont {Guo}},
  \bibinfo {author} {\bibfnamefont {S.}~\bibnamefont {Stramaglia}}, \ and\
  \bibinfo {author} {\bibfnamefont {F.~E.}\ \bibnamefont {Rosas}},\ }\href@noop
  {} {\bibfield  {journal} {\bibinfo  {journal} {Physical Review Research}\
  }\textbf {\bibinfo {volume} {4}},\ \bibinfo {pages} {013184} (\bibinfo {year}
  {2022})}\BibitemShut {NoStop}%
\bibitem [{\citenamefont {Hidalgo}\ \emph {et~al.}(2014)\citenamefont
  {Hidalgo}, \citenamefont {Grilli}, \citenamefont {Suweis}, \citenamefont
  {Munoz}, \citenamefont {Banavar},\ and\ \citenamefont
  {Maritan}}]{hidalgo2014information}%
  \BibitemOpen
  \bibfield  {author} {\bibinfo {author} {\bibfnamefont {J.}~\bibnamefont
  {Hidalgo}}, \bibinfo {author} {\bibfnamefont {J.}~\bibnamefont {Grilli}},
  \bibinfo {author} {\bibfnamefont {S.}~\bibnamefont {Suweis}}, \bibinfo
  {author} {\bibfnamefont {M.~A.}\ \bibnamefont {Munoz}}, \bibinfo {author}
  {\bibfnamefont {J.~R.}\ \bibnamefont {Banavar}}, \ and\ \bibinfo {author}
  {\bibfnamefont {A.}~\bibnamefont {Maritan}},\ }\href@noop {} {\bibfield
  {journal} {\bibinfo  {journal} {Proceedings of the National Academy of
  Sciences}\ }\textbf {\bibinfo {volume} {111}},\ \bibinfo {pages} {10095}
  (\bibinfo {year} {2014})}\BibitemShut {NoStop}%
\bibitem [{\citenamefont {Mora}\ and\ \citenamefont
  {Bialek}(2011)}]{mora2011biological}%
  \BibitemOpen
  \bibfield  {author} {\bibinfo {author} {\bibfnamefont {T.}~\bibnamefont
  {Mora}}\ and\ \bibinfo {author} {\bibfnamefont {W.}~\bibnamefont {Bialek}},\
  }\href@noop {} {\bibfield  {journal} {\bibinfo  {journal} {Journal of
  Statistical Physics}\ }\textbf {\bibinfo {volume} {144}},\ \bibinfo {pages}
  {268} (\bibinfo {year} {2011})}\BibitemShut {NoStop}%
\bibitem [{\citenamefont {Langton}(1990)}]{langton1990computation}%
  \BibitemOpen
  \bibfield  {author} {\bibinfo {author} {\bibfnamefont {C.~G.}\ \bibnamefont
  {Langton}},\ }\href@noop {} {\bibfield  {journal} {\bibinfo  {journal}
  {Physica D: Nonlinear Phenomena}\ }\textbf {\bibinfo {volume} {42}},\
  \bibinfo {pages} {12} (\bibinfo {year} {1990})}\BibitemShut {NoStop}%
\bibitem [{\citenamefont {Bohm}\ \emph {et~al.}(2026)\citenamefont {Bohm},
  \citenamefont {Adami},\ and\ \citenamefont {Hintze}}]{bohm2026redundancy}%
  \BibitemOpen
  \bibfield  {author} {\bibinfo {author} {\bibfnamefont {C.}~\bibnamefont
  {Bohm}}, \bibinfo {author} {\bibfnamefont {C.}~\bibnamefont {Adami}}, \ and\
  \bibinfo {author} {\bibfnamefont {A.}~\bibnamefont {Hintze}},\ }\href@noop {}
  {\bibfield  {journal} {\bibinfo  {journal} {Entropy}\ }\textbf {\bibinfo
  {volume} {28}},\ \bibinfo {pages} {167} (\bibinfo {year} {2026})}\BibitemShut
  {NoStop}%
\bibitem [{\citenamefont {Rosas}\ \emph {et~al.}(2019)\citenamefont {Rosas},
  \citenamefont {Mediano}, \citenamefont {Gastpar},\ and\ \citenamefont
  {Jensen}}]{rosas2019quantifying}%
  \BibitemOpen
  \bibfield  {author} {\bibinfo {author} {\bibfnamefont {F.~E.}\ \bibnamefont
  {Rosas}}, \bibinfo {author} {\bibfnamefont {P.~A.}\ \bibnamefont {Mediano}},
  \bibinfo {author} {\bibfnamefont {M.}~\bibnamefont {Gastpar}}, \ and\
  \bibinfo {author} {\bibfnamefont {H.~J.}\ \bibnamefont {Jensen}},\
  }\href@noop {} {\bibfield  {journal} {\bibinfo  {journal} {Physical Review
  E}\ }\textbf {\bibinfo {volume} {100}},\ \bibinfo {pages} {032305} (\bibinfo
  {year} {2019})}\BibitemShut {NoStop}%
\bibitem [{\citenamefont {Watanabe}(1960)}]{watanabe1960information}%
  \BibitemOpen
  \bibfield  {author} {\bibinfo {author} {\bibfnamefont {S.}~\bibnamefont
  {Watanabe}},\ }\href@noop {} {\bibfield  {journal} {\bibinfo  {journal} {IBM
  Journal of research and development}\ }\textbf {\bibinfo {volume} {4}},\
  \bibinfo {pages} {66} (\bibinfo {year} {1960})}\BibitemShut {NoStop}%
\bibitem [{\citenamefont {Garner}(1962)}]{garner1962uncertainty}%
  \BibitemOpen
  \bibfield  {author} {\bibinfo {author} {\bibfnamefont {W.~R.}\ \bibnamefont
  {Garner}},\ }\href@noop {} {\emph {\bibinfo {title} {Uncertainty and
  structure as psychological concepts.}}}\ (\bibinfo  {publisher} {Wiley},\
  \bibinfo {year} {1962})\BibitemShut {NoStop}%
\bibitem [{\citenamefont {Perez}(1977)}]{perez1977varepsilon}%
  \BibitemOpen
  \bibfield  {author} {\bibinfo {author} {\bibfnamefont {A.}~\bibnamefont
  {Perez}},\ }\href@noop {} {\bibfield  {journal} {\bibinfo  {journal}
  {Kybernetika}\ }\textbf {\bibinfo {volume} {13}},\ \bibinfo {pages} {439}
  (\bibinfo {year} {1977})}\BibitemShut {NoStop}%
\bibitem [{\citenamefont {Gatica}\ \emph {et~al.}(2022)\citenamefont {Gatica},
  \citenamefont {E.~Rosas}, \citenamefont {AM~Mediano}, \citenamefont {Diez},
  \citenamefont {P.~Swinnen}, \citenamefont {Orio}, \citenamefont {Cofr{\'e}},\
  and\ \citenamefont {M.~Cortes}}]{gatica2022high}%
  \BibitemOpen
  \bibfield  {author} {\bibinfo {author} {\bibfnamefont {M.}~\bibnamefont
  {Gatica}}, \bibinfo {author} {\bibfnamefont {F.}~\bibnamefont {E.~Rosas}},
  \bibinfo {author} {\bibfnamefont {P.}~\bibnamefont {AM~Mediano}}, \bibinfo
  {author} {\bibfnamefont {I.}~\bibnamefont {Diez}}, \bibinfo {author}
  {\bibfnamefont {S.}~\bibnamefont {P.~Swinnen}}, \bibinfo {author}
  {\bibfnamefont {P.}~\bibnamefont {Orio}}, \bibinfo {author} {\bibfnamefont
  {R.}~\bibnamefont {Cofr{\'e}}}, \ and\ \bibinfo {author} {\bibfnamefont
  {J.}~\bibnamefont {M.~Cortes}},\ }\href@noop {} {\bibfield  {journal}
  {\bibinfo  {journal} {PLoS Computational Biology}\ }\textbf {\bibinfo
  {volume} {18}},\ \bibinfo {pages} {e1010431} (\bibinfo {year}
  {2022})}\BibitemShut {NoStop}%
\bibitem [{\citenamefont {Faes}\ \emph {et~al.}(2022)\citenamefont {Faes},
  \citenamefont {Mijatovic}, \citenamefont {Antonacci}, \citenamefont
  {Pernice}, \citenamefont {Bar{\`a}}, \citenamefont {Sparacino}, \citenamefont
  {Sammartino}, \citenamefont {Porta}, \citenamefont {Marinazzo},\ and\
  \citenamefont {Stramaglia}}]{faes2022new}%
  \BibitemOpen
  \bibfield  {author} {\bibinfo {author} {\bibfnamefont {L.}~\bibnamefont
  {Faes}}, \bibinfo {author} {\bibfnamefont {G.}~\bibnamefont {Mijatovic}},
  \bibinfo {author} {\bibfnamefont {Y.}~\bibnamefont {Antonacci}}, \bibinfo
  {author} {\bibfnamefont {R.}~\bibnamefont {Pernice}}, \bibinfo {author}
  {\bibfnamefont {C.}~\bibnamefont {Bar{\`a}}}, \bibinfo {author}
  {\bibfnamefont {L.}~\bibnamefont {Sparacino}}, \bibinfo {author}
  {\bibfnamefont {M.}~\bibnamefont {Sammartino}}, \bibinfo {author}
  {\bibfnamefont {A.}~\bibnamefont {Porta}}, \bibinfo {author} {\bibfnamefont
  {D.}~\bibnamefont {Marinazzo}}, \ and\ \bibinfo {author} {\bibfnamefont
  {S.}~\bibnamefont {Stramaglia}},\ }\href@noop {} {\bibfield  {journal}
  {\bibinfo  {journal} {IEEE Transactions on Signal Processing}\ }\textbf
  {\bibinfo {volume} {70}},\ \bibinfo {pages} {5766} (\bibinfo {year}
  {2022})}\BibitemShut {NoStop}%
\bibitem [{\citenamefont {Bishop}(2006)}]{Bishop2006}%
  \BibitemOpen
  \bibfield  {author} {\bibinfo {author} {\bibfnamefont {C.~M.}\ \bibnamefont
  {Bishop}},\ }\href@noop {} {\emph {\bibinfo {title} {Pattern recognition and
  machine learning}}}\ (\bibinfo  {publisher} {springer},\ \bibinfo {year}
  {2006})\BibitemShut {NoStop}%
\bibitem [{\citenamefont {Barrett}(2015)}]{barrett2015exploration}%
  \BibitemOpen
  \bibfield  {author} {\bibinfo {author} {\bibfnamefont {A.~B.}\ \bibnamefont
  {Barrett}},\ }\href {\doibase 10.1103/PhysRevE.91.052802} {\bibfield
  {journal} {\bibinfo  {journal} {Phys. Rev. E}\ }\textbf {\bibinfo {volume}
  {91}},\ \bibinfo {pages} {052802} (\bibinfo {year} {2015})}\BibitemShut
  {NoStop}%
\bibitem [{\citenamefont {Koller}\ \emph {et~al.}(2009)\citenamefont {Koller},
  \citenamefont {Friedman},\ and\ \citenamefont
  {Bach}}]{koller2009probabilistic}%
  \BibitemOpen
  \bibfield  {author} {\bibinfo {author} {\bibfnamefont {D.}~\bibnamefont
  {Koller}}, \bibinfo {author} {\bibfnamefont {N.}~\bibnamefont {Friedman}}, \
  and\ \bibinfo {author} {\bibfnamefont {F.}~\bibnamefont {Bach}},\ }\href@noop
  {} {\emph {\bibinfo {title} {Probabilistic graphical models: principles and
  techniques}}}\ (\bibinfo  {publisher} {MIT press},\ \bibinfo {year}
  {2009})\BibitemShut {NoStop}%
\bibitem [{\citenamefont {Pearl}(1982)}]{pearl1982reverend}%
  \BibitemOpen
  \bibfield  {author} {\bibinfo {author} {\bibfnamefont {J.}~\bibnamefont
  {Pearl}},\ }in\ \href@noop {} {\emph {\bibinfo {booktitle} {Proceedings of
  the Second National Conference on Artificial Intelligence}}}\ (\bibinfo
  {publisher} {AAAI Press},\ \bibinfo {year} {1982})\ pp.\ \bibinfo {pages}
  {133--136}\BibitemShut {NoStop}%
\bibitem [{\citenamefont {Minka}(2001)}]{minka2001expectation}%
  \BibitemOpen
  \bibfield  {author} {\bibinfo {author} {\bibfnamefont {T.~P.}\ \bibnamefont
  {Minka}},\ }in\ \href@noop {} {\emph {\bibinfo {booktitle} {Proceedings of
  the Seventeenth conference on Uncertainty in artificial intelligence}}}\
  (\bibinfo {organization} {Morgan Kaufmann Publishers Inc.},\ \bibinfo {year}
  {2001})\ pp.\ \bibinfo {pages} {362--369}\BibitemShut {NoStop}%
\bibitem [{\citenamefont {Rosas}\ \emph {et~al.}(2024)\citenamefont {Rosas},
  \citenamefont {Mediano},\ and\ \citenamefont
  {Gastpar}}]{rosas2024characterising}%
  \BibitemOpen
  \bibfield  {author} {\bibinfo {author} {\bibfnamefont {F.~E.}\ \bibnamefont
  {Rosas}}, \bibinfo {author} {\bibfnamefont {P.~A.}\ \bibnamefont {Mediano}},
  \ and\ \bibinfo {author} {\bibfnamefont {M.}~\bibnamefont {Gastpar}},\ }in\
  \href@noop {} {\emph {\bibinfo {booktitle} {2024 IEEE International Symposium
  on Information Theory Workshops (ISIT-W)}}}\ (\bibinfo {organization}
  {IEEE},\ \bibinfo {year} {2024})\ pp.\ \bibinfo {pages} {1--6}\BibitemShut
  {NoStop}%
\bibitem [{\citenamefont {Han}(1975)}]{han1975linear}%
  \BibitemOpen
  \bibfield  {author} {\bibinfo {author} {\bibfnamefont {T.~S.}\ \bibnamefont
  {Han}},\ }\href@noop {} {\bibfield  {journal} {\bibinfo  {journal}
  {Information and Control}\ }\textbf {\bibinfo {volume} {29}},\ \bibinfo
  {pages} {337} (\bibinfo {year} {1975})}\BibitemShut {NoStop}%
\bibitem [{\citenamefont {Pearl}(2009)}]{pearl2009causal}%
  \BibitemOpen
  \bibfield  {author} {\bibinfo {author} {\bibfnamefont {J.}~\bibnamefont
  {Pearl}},\ }\href@noop {} {\bibfield  {journal} {\bibinfo  {journal}
  {Statistics Surveys}\ }\textbf {\bibinfo {volume} {3}},\ \bibinfo {pages}
  {96} (\bibinfo {year} {2009})}\BibitemShut {NoStop}%
\bibitem [{\citenamefont {Chickering}(1996)}]{chickering1996learning}%
  \BibitemOpen
  \bibfield  {author} {\bibinfo {author} {\bibfnamefont {D.~M.}\ \bibnamefont
  {Chickering}},\ }in\ \href@noop {} {\emph {\bibinfo {booktitle} {Learning
  from data: Artificial intelligence and statistics V}}}\ (\bibinfo
  {publisher} {Springer},\ \bibinfo {year} {1996})\ pp.\ \bibinfo {pages}
  {121--130}\BibitemShut {NoStop}%
\bibitem [{\citenamefont {Chow}\ and\ \citenamefont
  {Liu}(1968)}]{chow1968approximating}%
  \BibitemOpen
  \bibfield  {author} {\bibinfo {author} {\bibfnamefont {C.}~\bibnamefont
  {Chow}}\ and\ \bibinfo {author} {\bibfnamefont {C.}~\bibnamefont {Liu}},\
  }\href@noop {} {\bibfield  {journal} {\bibinfo  {journal} {IEEE transactions
  on Information Theory}\ }\textbf {\bibinfo {volume} {14}},\ \bibinfo {pages}
  {462} (\bibinfo {year} {1968})}\BibitemShut {NoStop}%
\bibitem [{\citenamefont {Berkson}(1946)}]{berkson1946}%
  \BibitemOpen
  \bibfield  {author} {\bibinfo {author} {\bibfnamefont {J.}~\bibnamefont
  {Berkson}},\ }\href@noop {} {\bibfield  {journal} {\bibinfo  {journal}
  {Biometrics Bulletin}\ }\textbf {\bibinfo {volume} {2}},\ \bibinfo {pages}
  {47} (\bibinfo {year} {1946})}\BibitemShut {NoStop}%
\bibitem [{\citenamefont {Munaf{\`o}}\ \emph {et~al.}(2018)\citenamefont
  {Munaf{\`o}}, \citenamefont {Tilling}, \citenamefont {Taylor}, \citenamefont
  {Evans},\ and\ \citenamefont {Davey~Smith}}]{munafo2018collider}%
  \BibitemOpen
  \bibfield  {author} {\bibinfo {author} {\bibfnamefont {M.~R.}\ \bibnamefont
  {Munaf{\`o}}}, \bibinfo {author} {\bibfnamefont {K.}~\bibnamefont {Tilling}},
  \bibinfo {author} {\bibfnamefont {A.~E.}\ \bibnamefont {Taylor}}, \bibinfo
  {author} {\bibfnamefont {D.~M.}\ \bibnamefont {Evans}}, \ and\ \bibinfo
  {author} {\bibfnamefont {G.}~\bibnamefont {Davey~Smith}},\ }\href@noop {}
  {\bibfield  {journal} {\bibinfo  {journal} {International journal of
  epidemiology}\ }\textbf {\bibinfo {volume} {47}},\ \bibinfo {pages} {226}
  (\bibinfo {year} {2018})}\BibitemShut {NoStop}%
\bibitem [{\citenamefont {Griffith}\ \emph {et~al.}(2020)\citenamefont
  {Griffith}, \citenamefont {Morris}, \citenamefont {Tudball}, \citenamefont
  {Herbert}, \citenamefont {Mancano}, \citenamefont {Pike}, \citenamefont
  {Sharp}, \citenamefont {Sterne}, \citenamefont {Palmer}, \citenamefont
  {Davey~Smith} \emph {et~al.}}]{griffith2020collider}%
  \BibitemOpen
  \bibfield  {author} {\bibinfo {author} {\bibfnamefont {G.~J.}\ \bibnamefont
  {Griffith}}, \bibinfo {author} {\bibfnamefont {T.~T.}\ \bibnamefont
  {Morris}}, \bibinfo {author} {\bibfnamefont {M.~J.}\ \bibnamefont {Tudball}},
  \bibinfo {author} {\bibfnamefont {A.}~\bibnamefont {Herbert}}, \bibinfo
  {author} {\bibfnamefont {G.}~\bibnamefont {Mancano}}, \bibinfo {author}
  {\bibfnamefont {L.}~\bibnamefont {Pike}}, \bibinfo {author} {\bibfnamefont
  {G.~C.}\ \bibnamefont {Sharp}}, \bibinfo {author} {\bibfnamefont
  {J.}~\bibnamefont {Sterne}}, \bibinfo {author} {\bibfnamefont {T.~M.}\
  \bibnamefont {Palmer}}, \bibinfo {author} {\bibfnamefont {G.}~\bibnamefont
  {Davey~Smith}},  \emph {et~al.},\ }\href@noop {} {\bibfield  {journal}
  {\bibinfo  {journal} {Nature communications}\ }\textbf {\bibinfo {volume}
  {11}},\ \bibinfo {pages} {5749} (\bibinfo {year} {2020})}\BibitemShut
  {NoStop}%
\bibitem [{Note1()}]{Note1}%
  \BibitemOpen
  \bibinfo {note} {DPGMMs can be informally understood as a mixture of Gaussian
  components with a flexible (potentially unbounded) number of
  components.}\BibitemShut {Stop}%
\bibitem [{\citenamefont {Neal}(2000)}]{neal2000markov}%
  \BibitemOpen
  \bibfield  {author} {\bibinfo {author} {\bibfnamefont {R.~M.}\ \bibnamefont
  {Neal}},\ }\href@noop {} {\bibfield  {journal} {\bibinfo  {journal} {Journal
  of Computational and Graphical Statistics}\ }\textbf {\bibinfo {volume}
  {9}},\ \bibinfo {pages} {249} (\bibinfo {year} {2000})}\BibitemShut {NoStop}%
\bibitem [{\citenamefont {Rosas}\ \emph {et~al.}(2016)\citenamefont {Rosas},
  \citenamefont {Ntranos}, \citenamefont {Ellison}, \citenamefont {Pollin},\
  and\ \citenamefont {Verhelst}}]{rosas2016understanding}%
  \BibitemOpen
  \bibfield  {author} {\bibinfo {author} {\bibfnamefont {F.}~\bibnamefont
  {Rosas}}, \bibinfo {author} {\bibfnamefont {V.}~\bibnamefont {Ntranos}},
  \bibinfo {author} {\bibfnamefont {C.~J.}\ \bibnamefont {Ellison}}, \bibinfo
  {author} {\bibfnamefont {S.}~\bibnamefont {Pollin}}, \ and\ \bibinfo {author}
  {\bibfnamefont {M.}~\bibnamefont {Verhelst}},\ }\href@noop {} {\bibfield
  {journal} {\bibinfo  {journal} {Entropy}\ }\textbf {\bibinfo {volume} {18}},\
  \bibinfo {pages} {38} (\bibinfo {year} {2016})}\BibitemShut {NoStop}%
\bibitem [{\citenamefont {Comrey}\ and\ \citenamefont
  {Lee}(2013)}]{comrey2013first}%
  \BibitemOpen
  \bibfield  {author} {\bibinfo {author} {\bibfnamefont {A.~L.}\ \bibnamefont
  {Comrey}}\ and\ \bibinfo {author} {\bibfnamefont {H.~B.}\ \bibnamefont
  {Lee}},\ }\href@noop {} {\emph {\bibinfo {title} {A first course in factor
  analysis}}}\ (\bibinfo  {publisher} {Psychology press},\ \bibinfo {year}
  {2013})\BibitemShut {NoStop}%
\bibitem [{\citenamefont {Chan}\ \emph {et~al.}(2017)\citenamefont {Chan},
  \citenamefont {Stumpf},\ and\ \citenamefont {Babtie}}]{chan2017gene}%
  \BibitemOpen
  \bibfield  {author} {\bibinfo {author} {\bibfnamefont {T.~E.}\ \bibnamefont
  {Chan}}, \bibinfo {author} {\bibfnamefont {M.~P.}\ \bibnamefont {Stumpf}}, \
  and\ \bibinfo {author} {\bibfnamefont {A.~C.}\ \bibnamefont {Babtie}},\
  }\href@noop {} {\bibfield  {journal} {\bibinfo  {journal} {Cell systems}\
  }\textbf {\bibinfo {volume} {5}},\ \bibinfo {pages} {251} (\bibinfo {year}
  {2017})}\BibitemShut {NoStop}%
\bibitem [{\citenamefont {Krakauer}\ and\ \citenamefont
  {Plotkin}(2002)}]{krakauer2002redundancy}%
  \BibitemOpen
  \bibfield  {author} {\bibinfo {author} {\bibfnamefont {D.~C.}\ \bibnamefont
  {Krakauer}}\ and\ \bibinfo {author} {\bibfnamefont {J.~B.}\ \bibnamefont
  {Plotkin}},\ }\href@noop {} {\bibfield  {journal} {\bibinfo  {journal}
  {Proceedings of the National Academy of Sciences}\ }\textbf {\bibinfo
  {volume} {99}},\ \bibinfo {pages} {1405} (\bibinfo {year}
  {2002})}\BibitemShut {NoStop}%
\bibitem [{\citenamefont {Battail}(2007)}]{battail2007information}%
  \BibitemOpen
  \bibfield  {author} {\bibinfo {author} {\bibfnamefont {G.}~\bibnamefont
  {Battail}},\ }\href@noop {} {\bibfield  {journal} {\bibinfo  {journal}
  {Introduction to biosemiotics}\ ,\ \bibinfo {pages} {299}} (\bibinfo {year}
  {2007})}\BibitemShut {NoStop}%
\bibitem [{\citenamefont {Kitano}(2007)}]{kitano2007towards}%
  \BibitemOpen
  \bibfield  {author} {\bibinfo {author} {\bibfnamefont {H.}~\bibnamefont
  {Kitano}},\ }\href@noop {} {\bibfield  {journal} {\bibinfo  {journal}
  {Molecular systems biology}\ }\textbf {\bibinfo {volume} {3}},\ \bibinfo
  {pages} {137} (\bibinfo {year} {2007})}\BibitemShut {NoStop}%
\bibitem [{\citenamefont {Jensen}(1998)}]{jensen1998self}%
  \BibitemOpen
  \bibfield  {author} {\bibinfo {author} {\bibfnamefont {H.~J.}\ \bibnamefont
  {Jensen}},\ }\href@noop {} {\emph {\bibinfo {title} {{Self-organized
  Criticality: Emergent Complex Behavior in Physical and Biological
  Systems}}}},\ Vol.~\bibinfo {volume} {10}\ (\bibinfo  {publisher} {Cambridge
  University Press},\ \bibinfo {year} {1998})\BibitemShut {NoStop}%
\bibitem [{\citenamefont {Liardi}\ \emph
  {et~al.}(2025{\natexlab{b}})\citenamefont {Liardi}, \citenamefont
  {Blackburne}, \citenamefont {Rajpal}, \citenamefont {Rosas},\ and\
  \citenamefont {Mediano}}]{liardi2025scalable}%
  \BibitemOpen
  \bibfield  {author} {\bibinfo {author} {\bibfnamefont {A.}~\bibnamefont
  {Liardi}}, \bibinfo {author} {\bibfnamefont {G.}~\bibnamefont {Blackburne}},
  \bibinfo {author} {\bibfnamefont {H.}~\bibnamefont {Rajpal}}, \bibinfo
  {author} {\bibfnamefont {F.~E.}\ \bibnamefont {Rosas}}, \ and\ \bibinfo
  {author} {\bibfnamefont {P.~A.}\ \bibnamefont {Mediano}},\ }\href@noop {}
  {\bibfield  {journal} {\bibinfo  {journal} {arXiv preprint arXiv:2506.18498}\
  } (\bibinfo {year} {2025}{\natexlab{b}})}\BibitemShut {NoStop}%
\bibitem [{\citenamefont {Ehrlich}\ \emph {et~al.}(2024)\citenamefont
  {Ehrlich}, \citenamefont {Schick-Poland}, \citenamefont {Makkeh},
  \citenamefont {Lanfermann}, \citenamefont {Wollstadt},\ and\ \citenamefont
  {Wibral}}]{ehrlich2024partial}%
  \BibitemOpen
  \bibfield  {author} {\bibinfo {author} {\bibfnamefont {D.~A.}\ \bibnamefont
  {Ehrlich}}, \bibinfo {author} {\bibfnamefont {K.}~\bibnamefont
  {Schick-Poland}}, \bibinfo {author} {\bibfnamefont {A.}~\bibnamefont
  {Makkeh}}, \bibinfo {author} {\bibfnamefont {F.}~\bibnamefont {Lanfermann}},
  \bibinfo {author} {\bibfnamefont {P.}~\bibnamefont {Wollstadt}}, \ and\
  \bibinfo {author} {\bibfnamefont {M.}~\bibnamefont {Wibral}},\ }\href@noop {}
  {\bibfield  {journal} {\bibinfo  {journal} {Physical Review E}\ }\textbf
  {\bibinfo {volume} {110}},\ \bibinfo {pages} {014115} (\bibinfo {year}
  {2024})}\BibitemShut {NoStop}%
\bibitem [{\citenamefont {Bounoua}\ \emph {et~al.}(2024)\citenamefont
  {Bounoua}, \citenamefont {Franzese},\ and\ \citenamefont
  {Michiardi}}]{bounoua2024somegai}%
  \BibitemOpen
  \bibfield  {author} {\bibinfo {author} {\bibfnamefont {M.}~\bibnamefont
  {Bounoua}}, \bibinfo {author} {\bibfnamefont {G.}~\bibnamefont {Franzese}}, \
  and\ \bibinfo {author} {\bibfnamefont {P.}~\bibnamefont {Michiardi}},\ }in\
  \href@noop {} {\emph {\bibinfo {booktitle} {International Conference on
  Machine Learning}}}\ (\bibinfo {year} {2024})\BibitemShut {NoStop}%
\bibitem [{\citenamefont {Young}\ \emph {et~al.}(2021)\citenamefont {Young},
  \citenamefont {Petri},\ and\ \citenamefont {Peixoto}}]{young2021hypergraph}%
  \BibitemOpen
  \bibfield  {author} {\bibinfo {author} {\bibfnamefont {J.-G.}\ \bibnamefont
  {Young}}, \bibinfo {author} {\bibfnamefont {G.}~\bibnamefont {Petri}}, \ and\
  \bibinfo {author} {\bibfnamefont {T.~P.}\ \bibnamefont {Peixoto}},\
  }\href@noop {} {\bibfield  {journal} {\bibinfo  {journal} {Communications
  Physics}\ }\textbf {\bibinfo {volume} {4}},\ \bibinfo {pages} {135} (\bibinfo
  {year} {2021})}\BibitemShut {NoStop}%
\bibitem [{\citenamefont {Malizia}\ \emph {et~al.}(2024)\citenamefont
  {Malizia}, \citenamefont {Corso}, \citenamefont {Gambuzza}, \citenamefont
  {Russo}, \citenamefont {Latora},\ and\ \citenamefont
  {Frasca}}]{malizia2024reconstructing}%
  \BibitemOpen
  \bibfield  {author} {\bibinfo {author} {\bibfnamefont {F.}~\bibnamefont
  {Malizia}}, \bibinfo {author} {\bibfnamefont {A.}~\bibnamefont {Corso}},
  \bibinfo {author} {\bibfnamefont {L.~V.}\ \bibnamefont {Gambuzza}}, \bibinfo
  {author} {\bibfnamefont {G.}~\bibnamefont {Russo}}, \bibinfo {author}
  {\bibfnamefont {V.}~\bibnamefont {Latora}}, \ and\ \bibinfo {author}
  {\bibfnamefont {M.}~\bibnamefont {Frasca}},\ }\href@noop {} {\bibfield
  {journal} {\bibinfo  {journal} {Nature Communications}\ }\textbf {\bibinfo
  {volume} {15}},\ \bibinfo {pages} {5184} (\bibinfo {year}
  {2024})}\BibitemShut {NoStop}%
\bibitem [{\citenamefont {Contisciani}\ \emph {et~al.}(2022)\citenamefont
  {Contisciani}, \citenamefont {Battiston},\ and\ \citenamefont
  {De~Bacco}}]{contisciani2022inference}%
  \BibitemOpen
  \bibfield  {author} {\bibinfo {author} {\bibfnamefont {M.}~\bibnamefont
  {Contisciani}}, \bibinfo {author} {\bibfnamefont {F.}~\bibnamefont
  {Battiston}}, \ and\ \bibinfo {author} {\bibfnamefont {C.}~\bibnamefont
  {De~Bacco}},\ }\href@noop {} {\bibfield  {journal} {\bibinfo  {journal}
  {Nature communications}\ }\textbf {\bibinfo {volume} {13}},\ \bibinfo {pages}
  {7229} (\bibinfo {year} {2022})}\BibitemShut {NoStop}%
\bibitem [{\citenamefont {Ganmor}\ \emph {et~al.}(2011)\citenamefont {Ganmor},
  \citenamefont {Segev},\ and\ \citenamefont {Schneidman}}]{ganmor2011sparse}%
  \BibitemOpen
  \bibfield  {author} {\bibinfo {author} {\bibfnamefont {E.}~\bibnamefont
  {Ganmor}}, \bibinfo {author} {\bibfnamefont {R.}~\bibnamefont {Segev}}, \
  and\ \bibinfo {author} {\bibfnamefont {E.}~\bibnamefont {Schneidman}},\
  }\href@noop {} {\bibfield  {journal} {\bibinfo  {journal} {Proceedings of the
  National Academy of sciences}\ }\textbf {\bibinfo {volume} {108}},\ \bibinfo
  {pages} {9679} (\bibinfo {year} {2011})}\BibitemShut {NoStop}%
\bibitem [{\citenamefont {Cayco-Gajic}\ \emph {et~al.}(2018)\citenamefont
  {Cayco-Gajic}, \citenamefont {Zylberberg},\ and\ \citenamefont
  {Shea-Brown}}]{cayco2018moment}%
  \BibitemOpen
  \bibfield  {author} {\bibinfo {author} {\bibfnamefont {N.~A.}\ \bibnamefont
  {Cayco-Gajic}}, \bibinfo {author} {\bibfnamefont {J.}~\bibnamefont
  {Zylberberg}}, \ and\ \bibinfo {author} {\bibfnamefont {E.}~\bibnamefont
  {Shea-Brown}},\ }\href@noop {} {\bibfield  {journal} {\bibinfo  {journal}
  {Entropy}\ }\textbf {\bibinfo {volume} {20}},\ \bibinfo {pages} {489}
  (\bibinfo {year} {2018})}\BibitemShut {NoStop}%
\bibitem [{\citenamefont {Decelle}\ \emph {et~al.}(2025)\citenamefont
  {Decelle}, \citenamefont {Navas~G{\'o}mez},\ and\ \citenamefont
  {Seoane}}]{decelle2025inferring}%
  \BibitemOpen
  \bibfield  {author} {\bibinfo {author} {\bibfnamefont {A.}~\bibnamefont
  {Decelle}}, \bibinfo {author} {\bibfnamefont {A.~d.~J.}\ \bibnamefont
  {Navas~G{\'o}mez}}, \ and\ \bibinfo {author} {\bibfnamefont {B.}~\bibnamefont
  {Seoane}},\ }\href@noop {} {\bibfield  {journal} {\bibinfo  {journal}
  {Physical Review Letters}\ }\textbf {\bibinfo {volume} {135}},\ \bibinfo
  {pages} {207301} (\bibinfo {year} {2025})}\BibitemShut {NoStop}%
\bibitem [{\citenamefont {Nemenman}\ \emph {et~al.}(2002)\citenamefont
  {Nemenman}, \citenamefont {Shafee},\ and\ \citenamefont
  {Bialek}}]{nemenman2002entropy}%
  \BibitemOpen
  \bibfield  {author} {\bibinfo {author} {\bibfnamefont {I.}~\bibnamefont
  {Nemenman}}, \bibinfo {author} {\bibfnamefont {F.}~\bibnamefont {Shafee}}, \
  and\ \bibinfo {author} {\bibfnamefont {W.}~\bibnamefont {Bialek}},\ }in\
  \href@noop {} {\emph {\bibinfo {booktitle} {Advances in neural information
  processing systems}}}\ (\bibinfo {year} {2002})\ pp.\ \bibinfo {pages}
  {471--478}\BibitemShut {NoStop}%
\bibitem [{\citenamefont {Gupta}\ and\ \citenamefont
  {Srivastava}(2010)}]{Gupta2010}%
  \BibitemOpen
  \bibfield  {author} {\bibinfo {author} {\bibfnamefont {M.}~\bibnamefont
  {Gupta}}\ and\ \bibinfo {author} {\bibfnamefont {S.}~\bibnamefont
  {Srivastava}},\ }\href {\doibase 10.3390/e12040818} {\bibfield  {journal}
  {\bibinfo  {journal} {Entropy}\ }\textbf {\bibinfo {volume} {12}},\ \bibinfo
  {pages} {818} (\bibinfo {year} {2010})}\BibitemShut {NoStop}%
\bibitem [{\citenamefont {Anderson}(2003)}]{Anderson2003}%
  \BibitemOpen
  \bibfield  {author} {\bibinfo {author} {\bibfnamefont {T.~W.}\ \bibnamefont
  {Anderson}},\ }\href@noop {} {\emph {\bibinfo {title} {An Introduction to
  Multivariate Statistical Analysis}}},\ \bibinfo {edition} {3rd}\ ed.\
  (\bibinfo  {publisher} {Wiley Interscience},\ \bibinfo {year}
  {2003})\BibitemShut {NoStop}%
\bibitem [{\citenamefont {Pav}(2015)}]{Pav2015}%
  \BibitemOpen
  \bibfield  {author} {\bibinfo {author} {\bibfnamefont {S.~E.}\ \bibnamefont
  {Pav}},\ }\href@noop {} {\bibfield  {journal} {\bibinfo  {journal} {arXiv
  preprint arXiv:1503.06266}\ } (\bibinfo {year} {2015})}\BibitemShut {NoStop}%
\bibitem [{\citenamefont {Siriteanu}\ \emph {et~al.}(2016)\citenamefont
  {Siriteanu}, \citenamefont {Kuriki}, \citenamefont {Richards},\ and\
  \citenamefont {Takemura}}]{Siriteanu2016}%
  \BibitemOpen
  \bibfield  {author} {\bibinfo {author} {\bibfnamefont {C.}~\bibnamefont
  {Siriteanu}}, \bibinfo {author} {\bibfnamefont {S.}~\bibnamefont {Kuriki}},
  \bibinfo {author} {\bibfnamefont {D.}~\bibnamefont {Richards}}, \ and\
  \bibinfo {author} {\bibfnamefont {A.}~\bibnamefont {Takemura}},\ }\href@noop
  {} {\bibfield  {journal} {\bibinfo  {journal} {Statistics \& Probability
  Letters}\ }\textbf {\bibinfo {volume} {115}},\ \bibinfo {pages} {79}
  (\bibinfo {year} {2016})}\BibitemShut {NoStop}%
\bibitem [{\citenamefont {Ng}\ and\ \citenamefont {Jordan}(2002)}]{Ng2001}%
  \BibitemOpen
  \bibfield  {author} {\bibinfo {author} {\bibfnamefont {A.~Y.}\ \bibnamefont
  {Ng}}\ and\ \bibinfo {author} {\bibfnamefont {M.~I.}\ \bibnamefont
  {Jordan}},\ }in\ \href@noop {} {\emph {\bibinfo {booktitle} {Advances in
  Neural Information Processing Systems 14}}},\ \bibinfo {editor} {edited by\
  \bibinfo {editor} {\bibfnamefont {T.~G.}\ \bibnamefont {Dietterich}},
  \bibinfo {editor} {\bibfnamefont {S.}~\bibnamefont {Becker}}, \ and\ \bibinfo
  {editor} {\bibfnamefont {Z.}~\bibnamefont {Ghahramani}}}\ (\bibinfo
  {publisher} {MIT Press},\ \bibinfo {year} {2002})\ pp.\ \bibinfo {pages}
  {841--848}\BibitemShut {NoStop}%
\bibitem [{\citenamefont {Kim}\ and\ \citenamefont {Pearl}(1983)}]{kim1983}%
  \BibitemOpen
  \bibfield  {author} {\bibinfo {author} {\bibfnamefont {J.}~\bibnamefont
  {Kim}}\ and\ \bibinfo {author} {\bibfnamefont {J.}~\bibnamefont {Pearl}},\
  }in\ \href@noop {} {\emph {\bibinfo {booktitle} {Proceedings IJCAI-83
  (Karlsruhe, Germany)}}}\ (\bibinfo  {publisher} {Morgan Kauffman},\ \bibinfo
  {address} {San Mateo, CA},\ \bibinfo {year} {1983})\ pp.\ \bibinfo {pages}
  {190--193}\BibitemShut {NoStop}%
\end{thebibliography}%

\clearpage
\newpage
\setcounter{secnumdepth}{3}

\onecolumngrid
\appendix

\section{The informational structure of pairwise Ising models}

\begin{figure*}[ht]
    \centering
    \includegraphics[width=\linewidth]{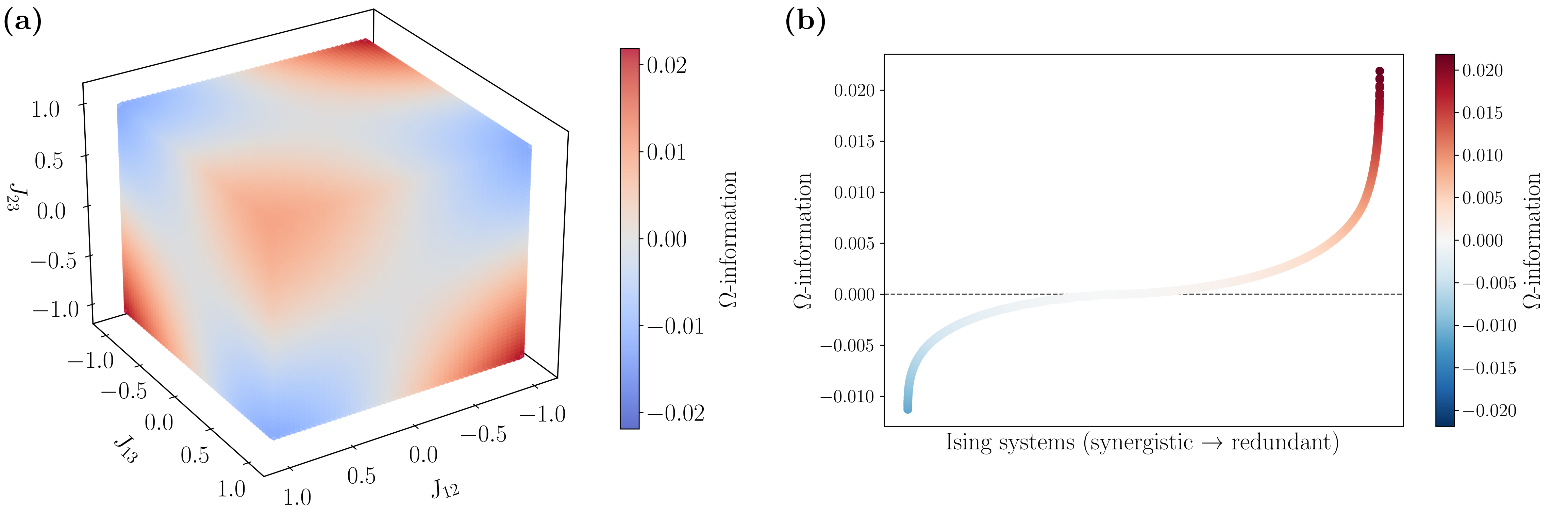}
    \caption{\textbf{The informational structure of the space of Ising models.} \textbf{(a)} O-information on the space of three-dimensional Ising models with zero external field and pairwise couplings $J_{ij}\in[-1,1], \,\,i,j=1,2,3$. \textbf{(b)} Ordered systems from maximally synergistic to maximally redundant. Predominantly synergistic systems compose 45\% of the space.}
    \label{fig:discrete_space}
\end{figure*}

Here, we investigate the discrete analogue of the Gaussian elliptope of Fig.~\ref{fig:fig0} by characterising the informational structure of pairwise maximum entropy models.
To achieve this, we consider the family of pairwise, zero-field Ising models on $n=3$ binary spins:
\begin{equation}
p(\bm{x}) \propto \exp\left(\sum_{i<j} J_{ij}x_ix_j\right) \,.
\end{equation}
Restricting the interactions to pairwise couplings isolates the three-dimensional parameter space $\{J_{12}, J_{13}, J_{23}\}$ within the full seven-dimensional simplex of joint distributions over the $2^3=8$ outcomes. Sweeping each coupling $J_{ij}$ independently over a regular grid, we then compute the probability distribution for each triplet and calculate its O-information. Unlike the Gaussian elliptope, whose domain is bounded by the positive-definiteness of the correlation matrix, every coupling triplet here defines a normalisable distribution. Consequently, the parameter space explored is a cube rather than a curved region (Fig.~\ref{fig:discrete_space}). 

As in the Gaussian case, redundancy dominates the unfrustrated regions corresponding to balanced signed graphs, whereas synergy is concentrated in frustrated configurations with unbalanced interactions. In contrast to the Gaussian setting, however, the parameter space is approximately evenly split between synergistic and redundant systems: 45\% of the models are predominantly synergistic and 55\% predominantly redundant.

\section{Loss of synergy in lower-order models} \label{app:synloss_sinfo}

In this section, we complement the results shown in Fig.~\ref{fig:synergyloss} by analysing the behaviour of S-information. Since the O-information only quantifies the balance between redundancy and synergy, we further investigate whether the increase in O-information observed in lower-order models arises from an increase in redundancy or a decrease in synergy.

To this end, we reproduce the analysis described in the main text. We sample distributions from either a Dirichlet or a DPGMM family (for the discrete and continuous settings, respectively) and fit their corresponding pairwise maximum entropy models, namely an Ising model and a Gaussian distribution. We then compute the S-information of both the original distribution, $\Sigma_{\text{Real}}$, and the fitted model, $\Sigma_{\text{Fit}}$.

\begin{figure}[h]
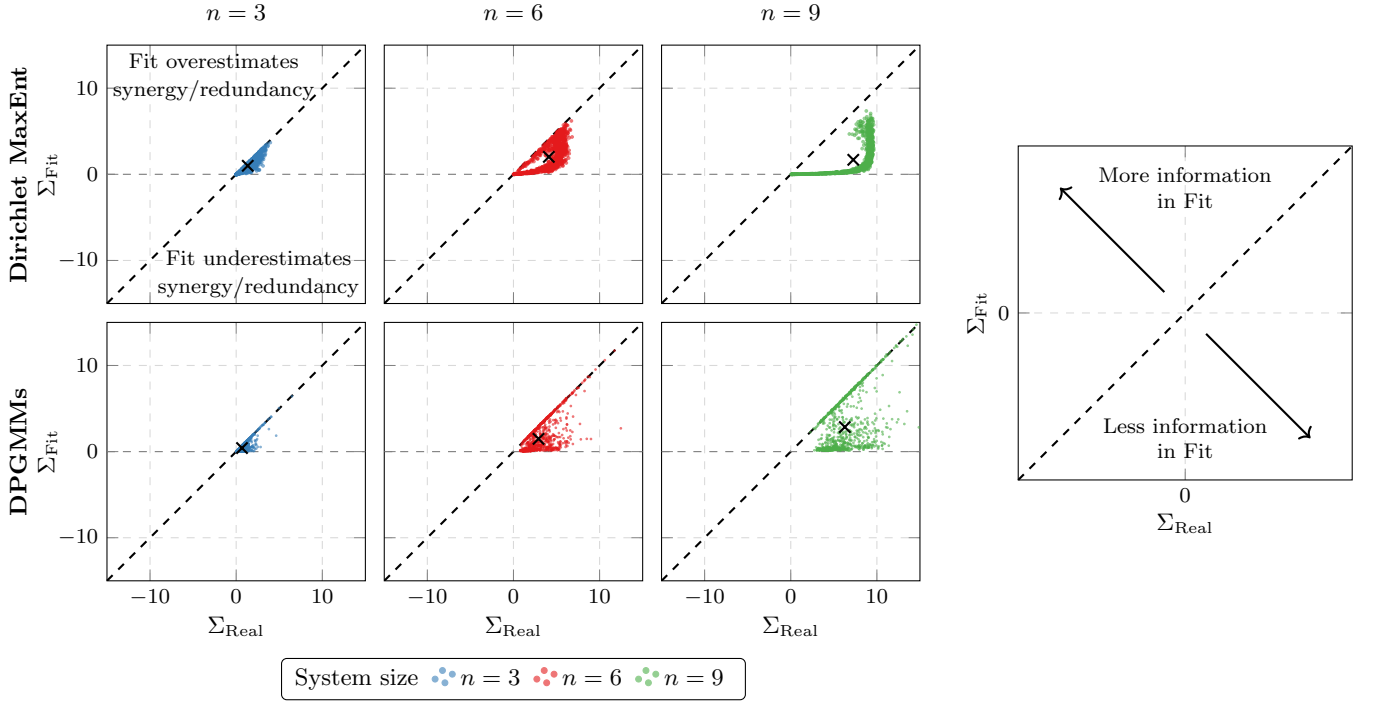

    \centering
    \includetikz{tikz/}{SynergyLoss_Sinfo_appendix}
    \caption{\textbf{S-information decreases in lower-order models}. (top) S-information of Dirichlet distributions and of their fit with a maximum entropy (Ising) distribution for $n=3, 6, 9$. (bottom) S-information of DPGMMs and of their fit with a multivariate Gaussian for $n=3, 6, 9$.}
    \label{fig:synloss_Sinfo}
\end{figure}

The results show that the S-information consistently decreases in the fitted models, indicating that these models overall contain less statistical information (Fig.~\ref{fig:synloss_Sinfo}). Combined with the increase in O-information reported in Fig.~\ref{fig:synergyloss}, this finding suggests that the information lost during the fitting procedure is predominantly synergistic, further supporting the hypothesis that lower-order models are largely insensitive to higher-order dependencies.

\section{Factor-analytic models underestimate synergy in simple Gaussian distributions} \label{app:FA_gaussian}

Here we accompany the factor-analytic analyses presented in the main text with an analogous investigation based on simple Gaussian models. Specifically, we sample covariance matrices from a Wishart distribution, compute the corresponding O- and S-information values ($\Omega_{\mathrm{Real}}$, $\Sigma_{\mathrm{Real}}$), and then fit factor models with varying numbers of factors, from which we again compute O- and S-information ($\Omega_{\mathrm{FA}}$, $\Sigma_{\mathrm{FA}}$). 

Unlike the procedure adopted in the main text, this approach avoids fitting a mixture of Gaussians to a single Gaussian distribution, which we previously identified as a source of synergy loss (Fig.~\ref{fig:synergyloss}). This allows us to isolate the effects of the factor-analytic approximation itself, ensuring that any observed behaviour can be attributed specifically to the factor-model fit.

\begin{figure}[h]
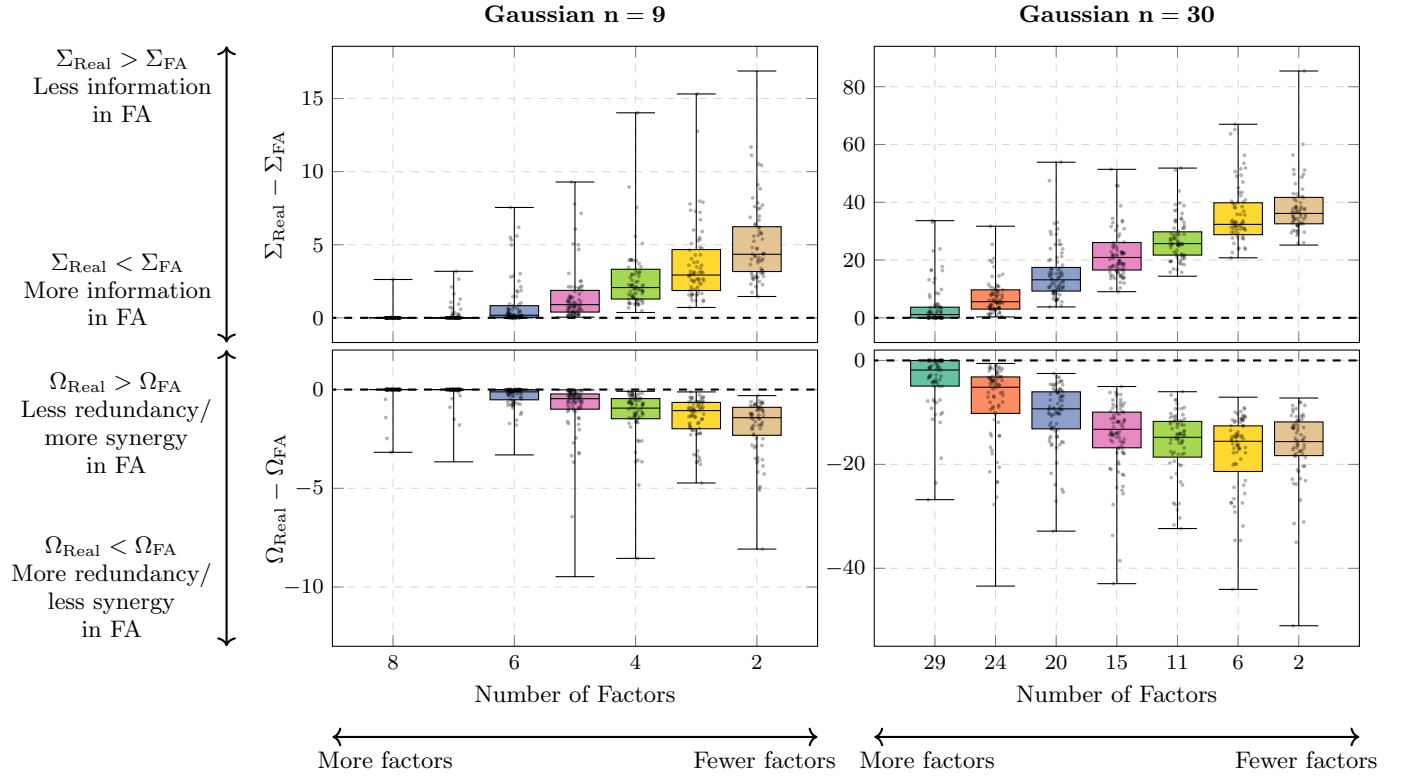

    \centering
    \includetikz{tikz/}{SynergyLoss_Gauss}
    \caption{\textbf{Factor analyses underestimate synergy in simple Gaussian distributions.}
    Difference between the $\Sigma$ (top) and $\Omega$ (bottom) information of Gaussians and their fit according to a given number of factors, for $n=9$ (left) and $n=30$ (right) variables. 
    }
    \label{fig:synergyloss_gauss}
\end{figure}

The results show that $\Sigma_{\mathrm{FA}}$ consistently decreases as the number of factors is reduced, indicating that both redundancy and synergy are attenuated in lower-order models (Fig.~\ref{fig:synergyloss_gauss}). At the same time, $\Omega_{\mathrm{FA}}$ increases for models with fewer factors, which may arise either from enhanced redundancy or from diminished synergy. Therefore, because the observed decrease in S-information implies a simultaneous reduction in both types of higher-order interdependence, these findings indicate that factor models systematically underestimate synergistic interactions.

\end{document}